\documentclass[a4paper,12pt]{article} \usepackage[english]{babel}
\usepackage{fullpage} \usepackage{epsfig} \usepackage{amsfonts}
\usepackage{amssymb} \usepackage{amstex} \usepackage{theorem}
\usepackage{eufrak}

\numberwithin{equation}{section}

\newtheorem{theorem}{Theorem}[section]
\newtheorem{corollary}[theorem]{Corollary}

\newtheorem{proposition}[theorem]{Proposition} {
\theorembodyfont{\normalfont}  } {
\theorembodyfont{\normalfont}  }
\makeatletter \title{Direction reversing travelling waves in the Fermi-Pasta-Ulam chain} \author{Bob Rink\thanks{Mathematics
Institute, Utrecht University,
  PO Box 80.010, 3508 TA Utrecht, The Netherlands, E-mail:
    {\tt rink@math.uu.nl} }} 
\begin{document} \maketitle
\abstract{\noindent This paper considers the famous
Fermi-Pasta-Ulam chain with periodic boundary
conditions and quartic nonlinearities. Due to special resonances and discrete symmetries, the
Birkhoff normal form of this Hamiltonian system is completely integrable, as was shown in \cite{Rink2}. We study how the level sets of the integrals foliate the phase space. Our study reveals all the integrable structure in the low energy domain of the chain. If the number of particles in the chain is even, then this foliation is singular. The method of singular reduction shows that the system has invariant pinched tori and monodromy. Monodromy is an obstruction to the existence of global action-angle
variables. The pinched tori are interpreted as homoclinic and heteroclinic
connections between travelling waves. Thus we discover a new class of solutions which can be described as direction reversing travelling waves. They remarkably show an interesting interaction of normal modes in the absense of energy transfer. These solutions can easily be observed numerically.}

\section{Introduction} \noindent The Fermi-Pasta-Ulam chain with
periodic boundary conditions is a model for point masses moving on
a circle with nonlinear forces acting between the nearest neighbours. The point masses constitute a one dimensional anharmonic lattice.\\ 
\indent This lattice is modelled as follows. Let $q_j \in \mathbb{R}$ be the position of the $j$-th
particle ($j=1,\ldots,n$) with respect to a certain reference position
on the circle. The space of positions $q=(q_1, \ldots, q_n)$ of the
particles in the chain is $\mathbb{R}^n$. The space of positions and
conjugate momenta is the cotangent bundle $T^*\mathbb{R}^n$ of
$\mathbb{R}^n$, the elements of which are denoted $(q, p) = (q_1,
\ldots, q_n, p_1, \ldots, p_n)$. $T^*\mathbb{R}^n$ is a symplectic
space, endowed with the canonical symplectic form $dq \wedge dp =
\sum_{j=1}^n dq_j \wedge dp_j$. Any smooth function $H: T^*\mathbb{R}^n
\to \mathbb{R}$ induces the Hamiltonian vector field $X_H$ given by
the defining relation $\iota_{X_H} (dq\wedge dp) = dH$. In other words,
we have the system of ordinary differential equations $\dot q_j =
\frac{\partial H}{\partial p_j}, \ \dot p_j = - \frac{\partial
H}{\partial q_j}$. \\
\indent The periodic FPU chain with quartic nonlinearities is the special
Hamiltonian system on $T^*\mathbb{R}^n$ induced by the
Hamiltonian

\begin{equation}\label{hamfpu} H = \sum_{j \in \mathbb{Z}/_{n
\mathbb{Z}}} \frac{1}{2} p_j^2 + W(q_{j+1} - q_j) \ , \end{equation} in
which $W: \mathbb{R} \to \mathbb{R}$ is a potential energy function of
the form \begin{equation} W(x) = \frac{1}{2} x^2 + \frac{1}{4} x^4 \ .
\end{equation} \noindent The system (\ref{hamfpu}) was introduced by E. Fermi, J. Pasta and S. Ulam 
\cite{Alamos}. These authors used the chain to model a string of which
the elements interact in a nonlinear way and they intended to study the ergodic properties of such a string. In their famous numerical
experiments, they surprisingly observed quasiperiodic behaviour and
no thermalisation or ergodicity. This puzzling phenomenon finally led to the
discovery of for instance solitons and stimulated the work on the KAM
theorem. In his recent book \cite{Weissert}, Weissert gives a rather
complete overview of the history of the `FPU problem' and its impact on
other branches of mathematics and physics. Ford's review article
\cite{Ford} contains a nice list of unsolved problems. \\
\indent Many different types of explanations were proposed for the observations of Fermi, Pasta and Ulam. Nishida \cite{Nishida} and Sanders \cite{Sanders} studied the Birkhoff normal form of Hamiltonian (\ref{hamfpu}). If this Birkhoff normal form were nondegenerately integrable, then the KAM theorem would prove that many low energy solutions of (\ref{hamfpu}) are quasiperiodic. These authors were forced to make very restrictive nonresonance assumptions though, so they never produced a proof.\\
\indent In \cite{Rink2} the Birkhoff normal form of the periodic FPU Hamiltonian (\ref{hamfpu}) was finally proved to be an integrable approximation of (\ref{hamfpu}). This remarkable result is caused by special resonances and discrete symmetries in the Hamiltonian (\ref{hamfpu}). The Liouville integrable Birkhoff normal form of (\ref{hamfpu}) is the subject of this paper. \\
\indent It turns out that the structure of the integrable approximation depends strongly on the parity of the number of particles $n$ in the chain. If $n$ is odd, then all the integrals of the Birkhoff normal form are quadratic functions of the phase space variables $(q, p)$. These quadratics form a set of global action variables. This implies that the foliation of the phase space in Liouville tori is trivial and that the equations of motion can be solved explicitly.\\
\indent The situation is not so simple though, if the number of particles $n$ is even. In this case one finds quadratic
as well as quartic integrals. They give rise to a singular foliation of the phase space. We shall find singular fibers that contain relative equilibria of focus-focus type. And the singular fibers on which they lie are pinched tori. These pinched tori are
homoclinic or heteroclinic connections between the focus-focus relative equilibria. We explicitly derive under what conditions these connections exist.\\
\indent In order to detect the pinched tori and the bifurcations of the relative equilibria, we use purely geometric arguments based on invariant theory and the method of singular reduction, see \cite{Cushman}. This geometric approach reveals all the integrable structure that is present in the Birkhoff normal form. It is for example well-known
\cite{Duistermaat}, \cite{Tienzung} that the presence of a pinched
torus results in nontrivial monodromy: the fibration of the phase space in Liouville tori is not trivial in the sense that the Liouville tori do not form a trivial torus bundle over the set of regular values of the integrals. Instead, we know how the Liouville tori are glued together globally, for instance on an energy level set. Nontrivial monodromy is an important obstruction to the
existence of global action-angle variables, see \cite{Duistermaat}. \\
\indent We expect that much of the integrable structure of the normal form is still present in the low energy domain of the original FPU chain (\ref{hamfpu}), meaning that the KAM tori are `glued together' in a similar way.\\
\indent On the other hand, our analysis yields interesting dynamical information. Previously mentioned relative equilibria can be interpreted as waves in the periodic FPU chain that travel clockwise and anti-clockwise. Thus it appears that there are homoclinic and heteroclinic connections between these travelling waves. In
the original system (\ref{hamfpu}), we indeed find these
direction reversing travelling waves numerically. They constitute a large collection of interesting new solutions of the periodic FPU chain. These solutions have the remarkable property that they show interesting interaction of normal modes without transferring energy. \\ 

\section{Phonons and reduction of the total momentum} \label{phonons}
\noindent The frequencies of the linear periodic FPU \ problem are:
\begin{equation}
\omega_j := 2 \sin( \frac{j \pi}{n} )  \ , \ \ j = 1, \ldots, n-1 \ . \end{equation} 
Note that $\omega_j = \omega_{n-j}$.\\ 
\indent We would like to view the solutions of the Hamiltonian system (\ref{hamfpu}) as a superposition of waves. Therefore we make the following Fourier
transformation:

\begin{equation} \label{matrix}
\hskip -1.3cm q_j = \sum_{1\leq k < \frac{n}{2}} \sqrt{\frac{2}{n\omega_k}} \left( \cos(\frac{2 \pi j k}{n}) \bar q_k  +  \sin(\frac{2 \pi j k}{n})\bar q_{n-k} \right) +  \frac{(-1)^j}{\sqrt{2n}} \bar q_{\frac{n}{n}} + \frac{1}{\sqrt{n}}\bar q_n 
\end{equation}

\noindent Of course, the term with the subscript $\frac{n}{2}$ only appears if $n$ is even.    \\
\indent The mapping $q \mapsto \bar q$ is a linear isomorphism of the position space $\mathbb{R}^n$. It induces a symplectic transformation $(q, p) \mapsto (\bar q, \bar p)$ on $T^*\mathbb{R}^n$.
The Fourier coefficients $(\bar q, \bar p)$ are known as {\it `phonons'} or {\it `quasi-particles'}.
In phonon-coordinates, the Hamiltonian (\ref{hamfpu}) reads
\begin{equation}\label{haminphononred}\label{haminphonon} H = \frac{1}{2}\bar p_n^2 +
\sum_{j=1}^{n-1} \frac{\omega_j}{2} (\bar p_j^2 + \bar q_j^2) + H_4(\bar
q_1, \ldots, \bar q_{n-1}) \  .  \end{equation}  $H_4$ denotes the quartic part
of $H$.  An exact expression for $H_4$ in terms of the $\bar q_j$ can be  computed by combining (\ref{hamfpu}) and
(\ref{matrix}). \\ 
\\ 
\noindent The pair of conjugate coordinates $(\bar q_j, \bar p_j)$ is called the $j$-th normal mode. For $1 \leq j \leq \frac{n}{2}$, we say that the $j$-th and the $(n-j)$-th normal mode have {\it wave number} $j$, because both are Fourier coefficients of waves with wave number $j$, that is waves with wave length $1/j$ times the lenght of the chain. The $j$-th and $(n-j)$-th normal mode also have the same linear vibrational frequency: $\omega_j = \omega_{n-j}$. But the $j$-th and $(n-j)$-th normal mode are out of phase, because one is described by a cosine and the other by a sine.\\
\\
\noindent Hamiltonian (\ref{haminphonon}) has an
obvious symmetry, namely the flow of $X_{\bar p_n} = \frac{\partial}{\partial \bar q_n}$. Clearly, this flow defines a free symplectic action on $T^*\mathbb{R}^n$ that leaves the Hamiltonian $H$ invariant.
The reduction of the symmetry is standard, cf. \cite{A&M} or \cite{Rink2}: by just forgetting about the constant $\frac{1}{2}\bar p_n^2$, $H$ trivially reduces to a Hamiltonian function of the variables $(\bar q_1, \ldots, \bar q_{n-1}, \bar p_1, \ldots, \bar p_{n-1}) \in T^*\mathbb{R}^{n-1}$. Thus we removed the total momentum from the equations of motion.\\
\indent Because $\omega_1,
\ldots, \omega_{n-1} > 0$, we conclude using Morse-lemma 
\cite{A&M} that the origin is a stable equilibrium of the Hamiltonian
system on $T^*\mathbb{R}^{n-1}$ induced by (\ref{haminphononred}). It corresponds to an
equidistant configuration of the $n$ particles in the chain. In the
sequel we will study the flow of the vector field induced by
(\ref{haminphononred}) on $T^*\mathbb{R}^{n-1}$ in a neighbourhood of this equilibrium.

\section{The Birkhoff normal form is integrable} We wish to simplify
(\ref{haminphononred}) by studying its average:  \begin{equation}\label{average}
\overline{H}(\bar q, \bar p) := \lim_{T \to \infty} \frac{1}{2T}
\int_{-T}^{T} H(e^{t X_{H_2}}(\bar q, \bar p)) \ dt \ .  \end{equation}
The flow of the linear vector field $X_{H_2}$ is easily calculated; it
is quasiperiodic with frequencies $\omega_j$. \\
\indent $\overline{H}$ is called a Birkhoff normal form for $H$.
Another convenient method to calculate normal forms, especially of
higher order, is the well-known Lie-series method, described for
instance in  \cite{churchillkummerrod} and \cite{Rink2}. In all cases,
one knows that there are canonical coordinates in a neighbourhood of
the origin of the phase space $T^*\mathbb{R}^{n-1}$, with the following
property: in these coordinates, $H$ looks like a higher
order perturbation of $\overline{H}$.  \\
\indent Thus, under certain nondegeneracy conditions, the low energy solutions of the Birkhoff normal form (\ref{average}) approximate the low energy solutions of the original system (\ref{haminphonon}). On the other hand, this is only true for finite $n$ and the domain where the approximation is valid probably shrinks when $n$ grows.
\\
\indent Before presenting $\overline{H}$,
let us introduce some notation that will be convenient later on. For $1 \leq j <\frac{n}{2}$, define the `Hopf variables'
\begin{equation}\label{abcd} \begin{array}{rl} u_j &\hskip-.6em := \bar q_j \bar
p_{n-j}  - \bar q_{n-j} \bar p_j \\ v_j &\hskip-.6em := \bar q_j \bar q_{n-j} +  \bar p_j \bar p_{n-j}  \\ w_j &\hskip-.6em := \frac{1}{2}(\bar q_j^2 + \bar
p_j^2 - \bar q_{n-j}^2 - \bar p_{n-j}^2) \end{array}
\end{equation} together with \begin{equation} \label{h123} 
\mathcal{H}_j := \frac{1}{2}(\bar q_j^2 + \bar p_j^2 + \bar q_{n-j}^2 +
\bar p_{n-j}^2) \ . \end{equation} 
\noindent And if $n$ is even
\begin{equation}
\mathcal{H}_{\frac{n}{2}} := \frac{1}{2}(\bar
q_{\frac{n}{2}}^2 + \bar p_{\frac{n}{2}}^2) \ .
\end{equation}
\noindent These quantities satisfy the
following relations:  \begin{equation} \label{relatiesabcd} u_j^2 +
v_j^2 + w_j^2 = \mathcal{H}_j^2 \ , \ \ 1 \leq j < \frac{n}{2} \ .  \end{equation} In terms of the Hopf variables, the following expression for $\overline{H}$ was found in \cite{Rink2}:
\begin{align}\label{normaalvorm} \nonumber
\overline{H} = \sum_{1\leq j\leq\frac{n}{2}} \omega_j \mathcal{H}_j + \frac{1}{n} \left\{ \sum_{1\leq k<l<\frac{n}{2}}
\frac{\omega_k \omega_l}{4} \mathcal{H}_k \mathcal{H}_l + \sum_{1 \leq k<\frac{n}{2}}
\frac{\omega_k^2}{32}  (3 \mathcal{H}_k^2 - u_k^2)  
+ \frac{1}{4} \mathcal{H}_{\frac{n}{2}}^2 + \right.  
\\
\left.
\frac{1}{2}\mathcal{H}_{\frac{n}{2}}\sum_{1 \leq k<\frac{n}{2}} \omega_k \mathcal{H}_k
 +\frac{1}{8} \sum_{1 \leq k < \frac{n}{4}} \omega_k \omega_{\frac{n}{2}-k}
 (v_k v_{\frac{n}{2}-k} -  w_k w_{\frac{n}{2}-k})
+\frac{1}{16} (v^2_{\frac{n}{4}} - w^2_{\frac{n}{4}})
\right\}  
\end{align}
\noindent In formula (\ref{normaalvorm}) it is understood that terms
with the subscript $\frac{n}{2}$ and $\frac{n}{4}$ only appear if $2$
respectively $4$ divides $n$.

%\begin{align}\label{normaalvorm} \nonumber \overline{H} = \mathcal{H}_1
%+ \sqrt{3} \mathcal{H}_2 &+ 2 \mathcal{H}_3 + \frac{3}{32}
%\mathcal{H}_1^2 + \frac{9}{32} \mathcal{H}_2^2 + \frac{1}{4}
%\mathcal{H}_3^2 + \frac{\sqrt{3}}{4} \mathcal{H}_1 \mathcal{H}_2  +
%\frac{1}{2} \mathcal{H}_3 \mathcal{H}_1  + \frac{\sqrt{3}}{2}
%\mathcal{H}_3 \mathcal{H}_2  \\ &- \frac{1}{32} v_1^2 - \frac{3}{32}
%w_1^2 + \frac{\sqrt{3}}{8} v_2 w_2 - \frac{\sqrt{3}}{8} v_3 w_3  \ .
%\end{align} 
As was shown in \cite{Rink2}, due to discrete symmetries and special eigenvalues of
Hamiltonian (\ref{hamfpu}), $\overline{H}$ remarkably is Liouville integrable.

\begin{proposition}\label{corr1}  If the number of particles $n$ is {\rm odd}, then the Birkhoff normal form 
(\ref{normaalvorm}) is Liouville
integrable with the quadratic 
integrals $\mathcal{H}_j, 
u_j$ $(1 \leq j \leq \frac{n-1}{2})$.
\end{proposition}

\noindent The integral $\mathcal{H}_j$ has the interpretation of the linear energy of the two modes with wave number $j$, whereas 
$u_j$ is the angular momentum of these modes.\\
\indent Since all these integrals induce a $2\pi$-periodic flow, they constitute a set of global action variables. The solutions of the normal form can explicitly be written down. And the foliation of the phase space into invariant tori is trivial in the sense that the set of Liouville tori is a trivial torus bundle over the set of regular values of the integrals. The normal form turns out to be even nondegenerate in the sense of the KAM theorem, which proves the abundance of quasiperiodic solutions in the low energy domain of the original system (\ref{hamfpu}). These statements were all proved in \cite{Rink2}. Thus, this system is quite tractable. 
\\
\indent But the situation is not so easy if the number of particles in the FPU chain is even.
\begin{proposition}\label{evenbeta}
If the number of particles $n$ is {\rm even}, then the Birkhoff normal form (\ref{normaalvorm}) is Liouville integrable. The integrals are the
quadratics 
$\mathcal{H}_j$ $(1 \leq j \leq \frac{n}{2})$,  $\mathcal{I}_j:=u_j -
u_{\frac{n}{2} - j}$ $(1 \leq j < \frac{n}{4})$ and $\mathcal{J}:=\frac{1}{2\sqrt{2n}} v_{\frac{n}{4}}$ (if
$n$ is a multiple of $4$) and the quartics $\mathcal{K}_j:=\frac{1}{32n} ( 4 \omega_j \omega_{\frac{n}{2}-j}
(v_j
v_{\frac{n}{2}-j} -  w_j
w_{\frac{n}{2}-j}) - \omega_j^2 u_j^2 -
\omega_{\frac{n}{2}-j}^2 u_{\frac{n}{2}-j}^2 )$ $(1 \leq j < \frac{n}{4})$.
\end{proposition} 
The rest of this paper shall be devoted to studying the dynamics and bifurcations of the integrable system (\ref{normaalvorm}) in the case that $n$ is even. We shall encounter nice phenomena such as whiskered tori with heteroclinic and homoclinic connections. They shall have the interpretation of direction reversing travelling waves and they constitute a class of solutions displaying an interesting interaction of Fourier modes with different wave numbers in the absence of energy transfer. \\
\indent We also study how the level sets of integrals in proposition \ref{evenbeta} foliate the phase space $T^*\mathbb{R}^{n-1}$. This study reveals all the integrable structure that is present in the low energy domain of the original FPU chain (\ref{hamfpu}).  It turns out that the foliation in Liouville tori is not trivial if $n \geq 6$ is even. There is monodromy and global action-angle variables can not exist. 

\section{Uncoupled equations and regular reduction} 
Let us use the short-hand notation $\tilde \jmath = \frac{n}{2}-j$.\\
\indent We want to study how the level sets of the integrals of proposition \ref{evenbeta} foliate the phase space $T^*\mathbb{R}^{n-1}$. Therefore it is very useful to note that these integrals are uncoupled in the following sense: the integral $\mathcal{H}_{\frac{n}{2}}$ depends only on the phase space variables $(\bar q_{\frac{n}{2}}, \bar p_{\frac{n}{2}})$. The integrals $\mathcal{H}_{\frac{n}{4}}$ and $\mathcal{J}$ depend only on the variables $(\bar q_{\frac{n}{4}}, \bar q_{\frac{3n}{4}}, \bar p_{\frac{n}{4}}, \bar p_{\frac{3n}{4}})$. The integrals $\mathcal{H}_j, \mathcal{H}_{\tilde \jmath}, \mathcal{I}_j$ and $\mathcal{K}_j$ depend only on the variables $(\bar q_j, \bar q_{\tilde \jmath},$ $\bar q_{n-j}, \bar q_{n-\tilde \jmath},$ $\bar p_j, \bar p_{\tilde \jmath},$ $\bar p_{n-j}, \bar p_{n-\tilde \jmath} )$. \\
\indent  Therefore it is sufficient to know how $\mathcal{H}_{\frac{n}{2}}$ foliates $T^*\mathbb{R}$, how $\mathcal{H}_{\frac{n}{4}}$ and $\mathcal{J}$ foliate  $T^*\mathbb{R}^2$ and how $\mathcal{H}_j, \mathcal{H}_{\tilde \jmath}, \mathcal{I}_j$ and $\mathcal{K}_j$ foliate $T^*\mathbb{R}^4$. The foliation of $T^*\mathbb{R}^{n-1}$ by all integrals is the cartesian product of these foliations. 
%\\ \indent Furthermore, $\overline{H} = f(\mathcal{H}_1, \ldots, \mathcal{H}_{\frac{n}{2}}) + \mathcal{J}^2 + \sum_j \mathcal{K}_j$ is the sum of Poisson commuting integrals. So the flow of $\overline{H}$ is simply the composition of the flows: $e^{t\overline{H}} = e^{t f(\mathcal{H})} \circ e^{t\mathcal{J}^2} \circ e^{t\mathcal{K}_1} \circ \ldots \circ  e^{t\mathcal{K}_?}$.
\subsection{The foliation of $T^*\mathbb{R}$}
\noindent It is clear that $\mathcal{H}_{\frac{n}{2}} = \frac{1}{2}(\bar
q_{\frac{n}{2}}^2 + \bar p_{\frac{n}{2}}^2)$ foliates $T^*\mathbb{R}$ in   circles which have their center at the origin. In other words, the Fourier mode with the highest wave number ($\frac{n}{2}$ waves in the chain) and the highest vibrational frequency ($\omega_{\frac{n}{2}} = 2$) has constant energy.
\subsection{The foliation of $T^*\mathbb{R}^2$}
The system on  $T^*\mathbb{R}^2$ with integrals $\mathcal{H}_{\frac{n}{4}}$ and $\mathcal{J}$ describes the interaction of the $\frac{n}{4}$-th and $\frac{3n}{4}$-th normal mode. These modes have equal wave number $\frac{n}{4}$ and equal vibrational frequency $\sqrt{2}$, but are out of phase. \\
\indent The foliation of $T^*\mathbb{R}^2$ can be described in the following standard way. It is clear that $\mbox{im}\mathcal{H}_{\frac{n}{4}} = \mathbb{R}_{\geq 0}$ and that for $h_{\frac{n}{4}} \in \mathbb{R}_{\geq 0}$, the inverse image $\mathcal{H}_{\frac{n}{4}}^{-1}( h_{\frac{n}{4}} ) = S^3_{\sqrt{2h_{\frac{n}{4}}}}$. Here $S^3_{\sqrt{2h_{\frac{n}{4}}}} \subset T^*\mathbb{R}^2$ is the three-dimensional sphere with radius $\sqrt{2h_{\frac{n}{4}}}$. The flow of $X_{\mathcal{H}_{\frac{n}{4}}}$ induces an $S^1$-symmetry on 
$S^3_{\sqrt{2h_{\frac{n}{4}}}}$ given by 

\begin{align} \label{werkingactieJ}
( \phi_{\frac{n}{4}}, \left( \begin{array}{l} \bar q_{\frac{n}{4}} \\  \bar q_{\frac{3n}{4}}\\ \bar p_{\frac{n}{4}}\\ \bar p_{\frac{3n}{4}} \end{array} \right) ) \mapsto \left( \begin{array}{l} \bar q_{\frac{n}{4}} \cos \phi_{\frac{n}{4}} + \bar p_{\frac{n}{4}} \sin\phi_{\frac{n}{4}} \\ 
\bar q_{\frac{3n}{4}} \cos\phi_{\frac{n}{4}} + \bar p_{\frac{3n}{4}} \sin\phi_{\frac{n}{4}}\\ \bar p_{\frac{n}{4}} \cos \phi_{\frac{n}{4}} - \bar q_{\frac{n}{4}} \sin\phi_{\frac{n}{4}}\\ \bar p_{\frac{3n}{4}} \cos\phi_{\frac{n}{4}} - \bar q_{\frac{3n}{4}} \sin\phi_{\frac{n}{4}} \end{array} \right)
\end{align}
The orbits of this circle action have dimension one if $h_{\frac{n}{4}} > 0$ and dimension zero otherwise. The reduced phase space $S^3_{\sqrt{2h_{\frac{n}{4}}}} /_{S^1}$ can be found by applying the Hopf map $\mathcal{F}^{(1)}:  S^3_{\sqrt{2h_{\frac{n}{4}}}} \to S^2_{h_{\frac{n}{4}} }$ which is defined by
$$\mathcal{F}^{(1)}: (\bar q, \bar p) \mapsto
(u_{\frac{n}{4}},v_{\frac{n}{4}},w_{\frac{n}{4}}) \ , $$  
see \cite{Cushman}. The fibers of
$\mathcal{F}^{(1)}$ are exactly the orbits of the $S^1$-action, so
$S^2_{h_{\frac{n}{4}} }$ constitutes the reduced phase space. Every
Hamiltonian function on $T^*\mathbb{R}^{2}$ that commutes with
$\mathcal{H}_{\frac{n}{4}}$ reduces to a
Hamiltonian on $S^2_{h_{\frac{n}{4}}}$ because it is constant on
the orbits of the $S^1$-action. In particular, $\mathcal{J} =\frac{1}{2 \sqrt{2n}} v_{\frac{n}{4}}$. The foliation of $S^2_{h_{\frac{n}{4}}}$ in level sets of $\mathcal{J}$ is trivial: the level sets are the circles of constant $v_{\frac{n}{4}}$. There are two stable relative equilibria. 
\begin{center} \epsfig{file=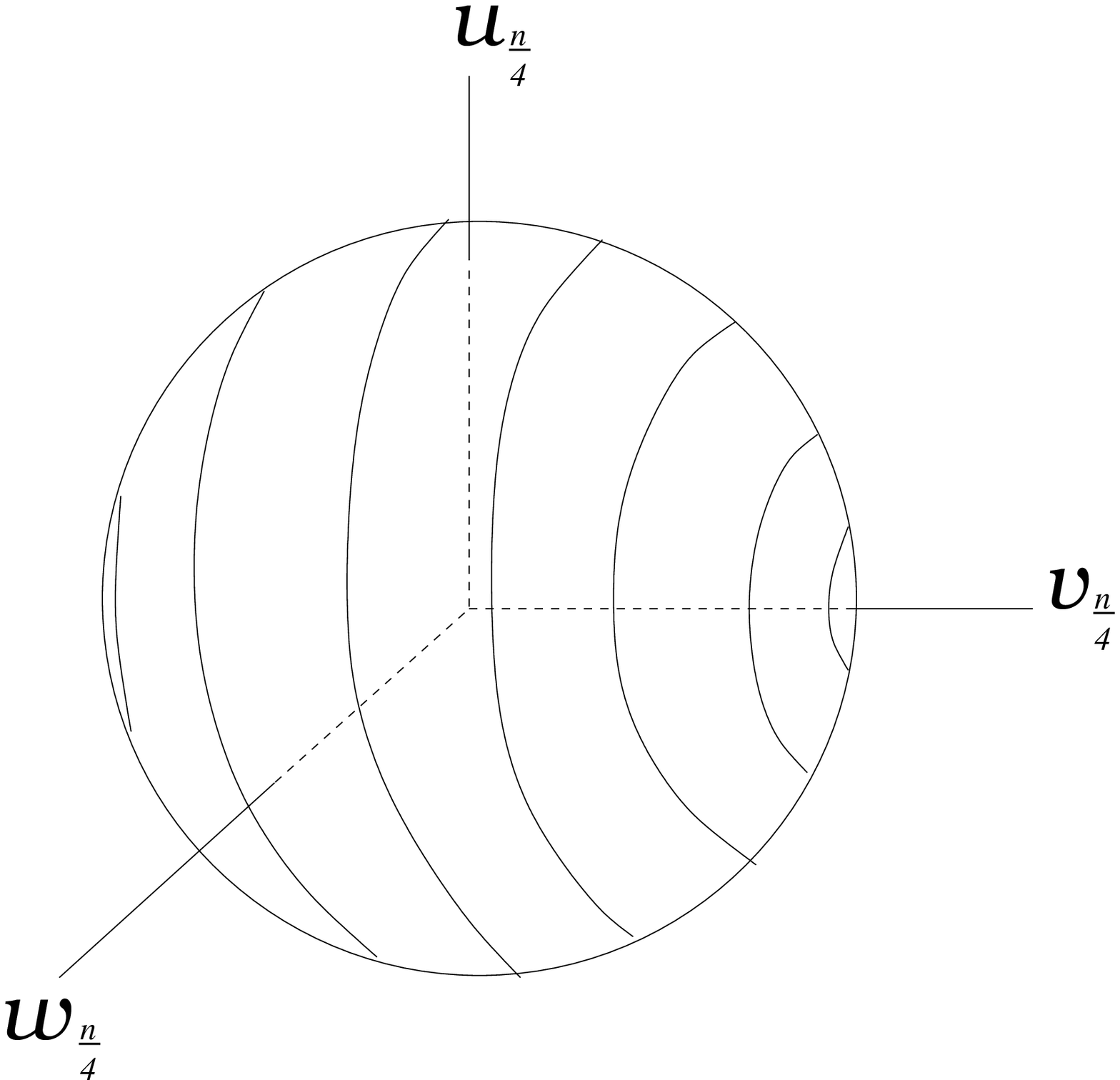 ,width=5cm ,height=5cm} \\
{\footnotesize The reduced space $S^2_{h_{\frac{n}{4}}}$ and the level sets of $\mathcal{J}$}
\end{center}
\noindent Reconstructing this picture by the Hopf map, we get the desired foliation of $T^*\mathbb{R}^2$.

\subsection{The foliation of $T^*\mathbb{R}^4$}
\noindent If $n \geq 6$ is even, then for each $1 \leq j < \frac{n}{4}$, we have the four commuting integrals $\mathcal{H}_j, \mathcal{H}_{\tilde \jmath}, \mathcal{I}_j$ and $\mathcal{K}_j$ on $T^*\mathbb{R}^4$. They describe the interaction between the two modes with wave number $j$ and the two modes with wave number $\tilde \jmath$. Note that these two pairs of modes do not interchange energy, since $\mathcal{H}_j$ and $\mathcal{H}_{\tilde \jmath}$ are constants. Still it turns out that their interaction is very interesting. The foliation of $T^*\mathbb{R}^4$ is difficult and contains singularities. \\
\indent On $T^*\mathbb{R}^4$, consider the
 commuting Hamiltonians $\mathcal{H}_j$ and $\mathcal{H}_{\tilde \jmath}$ defined in (\ref{h123}). They define a mapping
$\mathcal{H}_{j, \tilde \jmath}:=(\mathcal{H}_j, \mathcal{H}_{\tilde \jmath})$. It
is easy to see that $\mbox{im}\ \mathcal{H}_{j,\tilde \jmath} = (\mathbb{R}_{\geq
0})^{2}$, whereas for the level sets we know that
$\mathcal{H}_{j, \tilde \jmath}^{-1}(h_j, h_{\tilde \jmath}) = S^3_{\sqrt{2 h_j}}
\times   S^3_{\sqrt{2 h_{\tilde \jmath}}} $ modulo permutation
of coordinates. Here, $S^3_\varepsilon$ indicates the three-dimensional
sphere with radius $\varepsilon$. Note that $\varepsilon=0$ is not
excluded.\\ \indent Because the flows of the Hamiltonian vector fields
of $\mathcal{H}_j$ and $\mathcal{H}_{\tilde \jmath}$ commute
and are periodic with period $2 \pi$, they define a linear symplectic
$T^{2}$-action on $S^3_{\sqrt{2 h_j}}
\times S^3_{\sqrt{2 h_{\tilde \jmath}}}$ given by
\begin{align} \label{werkingactie}
( \left( \begin{array}{l} \phi_j \\ \phi_{\tilde \jmath} \end{array} \right), \left( \begin{array}{l} \bar q_j\\ \bar q_{\tilde \jmath}\\ \bar q_{n-\tilde \jmath}\\ \bar q_{n-j}\\ \bar p_j\\ \bar p_{\tilde \jmath}\\ \bar p_{n-\tilde \jmath}\\ \bar p_{n-j} \end{array} \right) ) \mapsto \left( \begin{array}{l} \bar q_j \cos \phi_j + \bar p_j \sin\phi_j\\ \bar q_{\tilde \jmath} \cos\phi_{\tilde \jmath} + \bar p_{\tilde \jmath} \sin\phi_{\tilde \jmath}\\ \bar q_{n-\tilde \jmath} \cos \phi_{\tilde \jmath} + \bar p_{n-\tilde \jmath} \sin\phi_{\tilde \jmath}\\ \bar q_{n-j} \cos\phi_{j} + \bar p_{n-j} \sin\phi_{j}\\ \bar p_j \cos  \phi_j - \bar q_j \sin\phi_j\\ \bar p_{\tilde \jmath} \cos\phi_{\tilde \jmath} - \bar p_{\tilde \jmath} \sin\phi_{\tilde \jmath}\\ \bar p_{n-\tilde \jmath} \cos \phi_{\tilde \jmath} - \bar q_{n-\tilde \jmath} \sin\phi_{\tilde \jmath}\\ \bar p_{n-j} \cos\phi_{j} - \bar q_{n-j} \sin\phi_{j} \end{array} \right)  
\end{align}
%\begin{equation} \nonumber
%\label{werkingactie}
% \left( \begin{array}{c} \phi_1 \\ \phi_2 \\ \phi_3 \end{array} \right)
% \times \left( \begin{array}{c} \bar q_1  \\ \bar q_2 \\ \bar q_3
% \\ \bar q_4 \\ \bar q_5 \\ \bar p_1 \\ \bar p_2 \\ \bar p_3 \\ \bar
% p_4 \\ \bar p_5 \end{array} \right) \mapsto \left( \begin{array}{c}
% \bar q_1 \cos \phi_1 + \bar p_1 \sin \phi_1 \\ \bar q_2 \cos \phi_1 +
% \bar p_2 \sin \phi_1 \\ \bar q_3 \cos \phi_2 + \bar p_3 \sin \phi_2
% \\ \bar q_4 \cos \phi_2 + \bar p_4 \sin \phi_2 \\ \bar q_5 \cos \phi_3
% + \bar p_5 \sin \phi_3 \\ \bar p_1 \cos \phi_1 - \bar q_1 \sin \phi_1
% \\ \bar p_2 \cos \phi_1 - \bar q_2 \sin \phi_1 \\ \bar p_3 \cos
% \phi_2  - \bar q_3 \sin \phi_2 \\ \bar p_4 \cos \phi_2 - \bar q_4 \sin
% \phi_2 \\ \bar p_5 \cos \phi_3 - \bar q_5 \sin \phi_3 \end{array}
% \right) \ .  \end{equation} 
\noindent The orbits of this $T^{2}$-action are all tori of dimension
$\#\{k\in\{j , \tilde \jmath\}\ |\ h_k \neq 0\}$. \\ \indent We want to study the
reduced phase-space $(S^3_{\sqrt{2 h_j}}
\times S^3_{\sqrt{2 h_{\tilde \jmath}}} ) /_{T^{2}}$. We do this by applying a second
reduction map $\mathcal{F}^{(2)}: S^3_{\sqrt{2 h_j}}
\times S^3_{\sqrt{2 h_{\tilde \jmath}}} \to S^2_{h_j} \times S^2_{h_{\tilde \jmath}}$ defined by $$\mathcal{F}^{(2)}: (\bar q, \bar p) \mapsto
(u_j,v_j,w_j, u_{\tilde \jmath}, v_{\tilde \jmath}, w_{\tilde \jmath}) \ . $$ \noindent $\mathcal{F}^{(2)}$ is the cartesian product of two Hopf mappings. The fibers of
$\mathcal{F}^{(2)}$ are exactly the orbits of the $T^{2}$-action, so
$S^2_{h_j} \times 
S^2_{h_{\tilde \jmath}}$ constitutes the reduced phase space. Every
Hamiltonian function on $T^*\mathbb{R}^{4}$ that commutes with
$\mathcal{H}_j$ and $\mathcal{H}_{\tilde \jmath}$ reduces to a
Hamiltonian on $S^2_{h_j} \times 
S^2_{h_{\tilde \jmath}}$ because it is constant on
the orbits of the $T^{2}$-action. In particular $\mathcal{I}_j$ and $\mathcal{K}_j$. The Hamiltonian
equations of motion induced by such a Hamiltonian reduce to Hamiltonian
equations on the orbit space $S^2_{h_j} \times S^2_{h_{\tilde \jmath}}$. For a Hamiltonian function $H$, these
reduced equations read \begin{equation}\nonumber \frac{d}{dt} \left(\begin{array}{l} u_k \\ v_k \\ w_k  \end{array}\right) = 2 \left( \begin{array}{l} u_k \\ v_k \\ w_k \end{array} \right)  \times \left( \begin{array}{l} \partial_{u_k} H \\ \partial_{v_k} H \\ \partial_{w_k} H \end{array} \right)    \ ,
\end{equation}
for $k = j, \tilde \jmath$. Thus we obtain a two degree of freedom integrable Hamiltonian system on the product of two spheres. We will now study this integrable system, which describes the interaction of waves with wave numbers $j$ and $\tilde \jmath$.

\section{Travelling waves} \label{trw} 
Consider the integrable system on $S^2_{h_{j}} \times S^2_{h_{\tilde \jmath }}$ with  Hamiltonian  $\mathcal{K}_j$ and momentum $\mathcal{I}_j = u_j - u_{\tilde \jmath}$. 
The flow
of  $X_{\mathcal{I}_j}$ induces  a symplectic $S^1$-action on $S^2_{h_{j}} \times S^2_{h_{\tilde \jmath }}$ given by
\begin{align} \label{s1actie} 
( t, \left( \begin{array}{l} u_j\\ v_j\\ w_j\\ u_{\tilde \jmath }\\ v_{\tilde \jmath }\\ w_{\tilde \jmath }\end{array} \right) ) \mapsto \left( \begin{array}{l}  u_j\\ v_j \cos
2t  +  w_j \sin 2t \\ w_j \cos 2t  -  v_j \sin 2t\\ u_{\tilde \jmath }\\ v_{\tilde \jmath } \cos
2t - w_{\tilde \jmath } \sin 2t \\ w_{\tilde \jmath }\cos 2t  +  v_{\tilde \jmath } \sin 2t \end{array} \right) 
\end{align}
This action has four isolated fixed points, namely the points $(\pm h_j, 0, 0,  \pm h_{\tilde \jmath }, 0, 0)$. But because the Hamiltonian $\mathcal{K}_j$ is invariant under the action (\ref{s1actie}), this implies that the derivative of $\mathcal{K}_j$ also vanishes at these points. In other words, the points $(\pm h_j, 0, 0,  \pm h_{\tilde \jmath }, 0, 0)$ constitute the
set of joint critical points of $\mathcal{I}_j$ and $\mathcal{K}_j$.\\ 
\indent Critical points of the reduced system on $S^2_{h_{j}} \times S^2_{h_{\tilde \jmath }}$ are called relative
equilibria, because in the reconstructed system on
$T^*\mathbb{R}^{4}$ - or $T^*\mathbb{R}^{n-1}$ if you like - their  fibers represent invariant sets, see \cite{A&M}. It follows
from (\ref{werkingactie}) that in $T^*\mathbb{R}^{4}$ the critical fiber
$(\mathcal{F}^{(2)})^{-1}(\pm_j h_j, 0, 0,  \pm_{\tilde \jmath} h_{\tilde \jmath }, 0, 0)$ is the following parametrised torus 
\begin{align}
& \{ \ (\sqrt{h_j}\cos \phi_j, \sqrt{h_{\tilde \jmath}}\cos \phi_{\tilde \jmath}, \sqrt{h_{\tilde \jmath}}\sin \phi_{\tilde \jmath}, \sqrt{h_j}\sin \phi_j, \\ \nonumber \mp_j \sqrt{h_j}\sin \phi_j&, \mp_{\tilde \jmath} \sqrt{h_{\tilde \jmath}}\sin \phi_{\tilde \jmath}, \pm_{\tilde \jmath} \sqrt{h_{\tilde \jmath}}\cos \phi_{\tilde \jmath}, \pm_j \sqrt{h_j}\cos \phi_j) \ |\ (\phi_j , \phi_{\tilde \jmath}) \in T^2\ \}
\end{align} 
It has dimension
$\#\{k\in\{j, \tilde \jmath  \}|h_k \neq 0\}$. This torus is invariant under the
flow of $X_{\overline{H}}$ so one can write the equations induced by
$\overline{H}$ as equations for $\phi \in T^{2}$. Using
expression (\ref{normaalvorm}) it is not hard to compute that they
read:  \begin{align} &\frac{d\phi_j}{dt} =  \pm_j ( \frac{\partial \overline{H}}{\partial \mathcal{H}_j} \left|_{\mathcal{H} = h} \right. - \frac{\omega_j^2 h_j}{16} ) \\
 &\frac{d\phi_{\tilde \jmath }}{dt} = \pm_{\tilde \jmath} ( \frac{\partial \overline{H}}{\partial \mathcal{H}_{\tilde \jmath}} \left|_{\mathcal{H} = h} \right. - \frac{\omega_{\tilde \jmath}^2 h_{\tilde \jmath}}{16} )
\end{align} 
Hence the
motion in the critical fibers is uniform. The corresponding solutions
have a clear physical interpretation: with (\ref{matrix}) one can make
the transformation back to the original coordinates: 
 \begin{align} \nonumber &q_k = \sqrt{\frac{2h_j}{n\omega_j}} \cos(\frac{2 \pi j k}{n}-\phi_j) +
\sqrt{\frac{2 h_{\tilde \jmath}}{n \omega_{\tilde \jmath}}} \cos(\frac{2 \pi \tilde \jmath k}{n}-\phi_{\tilde \jmath})\ . 
\end{align} 
\noindent In
other words, a solution that lies in a critical fiber is a
superposition of
\begin{itemize}
\item[1.] a travelling wave
$q_k = \cos(\frac{2 \pi j k}{n})$  with amplitude  $\sqrt{\frac{2 h_j}{n \omega_j}}$ and
speed  approximately $\pm_j \omega_j$.
\item[2.] a travelling wave
$q_k = \cos(\frac{2 \pi \tilde \jmath k}{n})$  with amplitude  $\sqrt{\frac{2 h_{\tilde \jmath}}{n \omega_{\tilde \jmath}}}$ and
speed  approximately $\pm_{\tilde \jmath} \omega_{\tilde \jmath}$.
\end{itemize}

\noindent The constants $h_j$ and $h_{\tilde \jmath}$
are supposed to be small, since otherwise the normal form approximation
has no validity. Thus such a solution is a superposition
of two travelling waves with wave number $j$ and $\tilde \jmath$. If $\pm_j=\pm_{\tilde \jmath }$ then these waves move in the
same direction. Otherwise one moves clockwise and the other moves 
anti-clockwise.\\ \indent Travelling waves and superposed travelling
waves have previously been studied in infinite FPU chains 
\cite{Iooss}. But they can obviously also occur in finite periodic
chains. We shall study the stability of the travelling wave solutions and homoclinic and heteroclinic connections
between them. We do this of course in the reduced context,
that is we consider them as critical points in the reduced phase space $S^2_{h_{j}} \times S^2_{h_{\tilde \jmath }}$.

\section{Stability of the relative equilibria} We want to determine the
stability type of the superposed travelling wave solutions in the Birkhoff normal form, that is the
stability type of the relative equilibria $(\pm h_j, 0, 0,$ $\pm h_{\tilde \jmath }, 0, 0)$.  For this purpose we introduce local coordinates on $S^2_{h_1}
\times S^2_{h_2}$ near $(\pm h_j, 0, 0,$ $\pm h_{\tilde \jmath }, 0, 0)$ by projecting
$(u_j, v_j, w_j, u_{\tilde \jmath}, v_{\tilde \jmath}, w_{\tilde \jmath}) \mapsto (v_j, w_j,  v_{\tilde \jmath}, w_{\tilde \jmath})$. Note that these are not Darboux
coordinates. The critical points themselves are all mapped to $(0, 0,
0, 0)$.
\subsection{A Lyapunov function} One way of proving stability is by pointing out a
Lyapunov function. The Hamiltonian $\mathcal{K}_j$ is the first
candidate since it is an a priori constant of motion. But it turns out
that $\mathcal{K}_j$ is not definite at any of the relative equilibria.
Luckily, we have another constant of motion, namely $\mathcal{I}_j$. We
now use that at $(\pm_j h_j, 0, 0,  \pm_{\tilde \jmath} h_{\tilde \jmath }, 0, 0)$ one may write $u_j
= \pm_j \sqrt{h_j^2 - v_j^2 - w_j^2} = \pm_j (h_j - \frac{1}{2h_j}(v_j^2
+ w_j^2) + \ldots)$ and $u_{\tilde \jmath}
= \pm_{\tilde \jmath} \sqrt{h_{\tilde \jmath}^2 - v_{\tilde \jmath}^2 - w_{\tilde \jmath}^2} = \pm_{\tilde \jmath} (h_{\tilde \jmath} - \frac{1}{2h_{\tilde \jmath}}(v_{\tilde \jmath}^2
+ w_{\tilde \jmath}^2) + \ldots)$. So $$\mathcal{I}_j =
u_j - u_{\tilde \jmath} = \pm_j h_j \mp_{\tilde \jmath} h_{\tilde \jmath} \mp_j  \frac{1}{2h_j}(v_j^2 + w_j^2)
\pm_{\tilde \jmath} \frac{1}{2h_{\tilde \jmath}}(v_{\tilde \jmath}^2 + w_{\tilde \jmath}^2) + \ldots$$ which is definite at $(0,
0 ,0, 0)$ if and only if $\pm_j \neq \pm_{\tilde \jmath}$. We conclude that the
relative equilibria $\pm (h_j, 0, 0,$ $-h_{\tilde \jmath }, 0, 0)$ are stable. \\
\indent Near the relative equilibria $\pm (h_j, 0, 0, h_{\tilde \jmath }, 0, 0)$ we will try to make linear combinations of $\mathcal{K}_j$ and $\mathcal{I}_j$ that are definite. It is easily computed that in local coordinates $(v_j, w_j, v_{\tilde \jmath}, w_{\tilde \jmath})$
$$(32 n \mathcal{K}_j \pm 2 \lambda h_j h_{\tilde \jmath} \mathcal{I}_j) = (\omega_j^2 + \lambda h_{\tilde \jmath})(v_j^2 + w_j^2) + (\omega_{\tilde \jmath}^2 - \lambda h_j)(v_{\tilde \jmath}^2 + w_{\tilde \jmath}^2) + 4 \omega_j \omega_{\tilde \jmath} (v_j v_{\tilde \jmath} - w_j w_{\tilde \jmath}) + \ldots$$
modulo constants. Using that $|v_j v_{\tilde \jmath} - w_j w_{\tilde \jmath}| \leq ||(v_j, w_j)||\cdot ||(v_{\tilde \jmath}, w_{\tilde \jmath})||$, one sees that this expression is definite if and only if  
$$\det \left( \begin{array}{cc} \omega_j^2 + \lambda h_{\tilde \jmath} & 2\omega_j \omega_{\tilde \jmath}\\ 2\omega_j \omega_{\tilde \jmath} & \omega_{\tilde \jmath}^2 - \lambda h_j \end{array} \right) = -\lambda^2 h_j h_{\tilde \jmath} + \lambda( \omega_{\tilde \jmath}^2 h_{\tilde \jmath} - \omega_j^2 h_j) - 3 \omega_j^2 \omega_{\tilde \jmath}^2 > 0 \ .
$$
The preceding inequality has real solutions $\lambda$ if the discriminant 
\begin{equation}\label{r}
r:= \omega_j^4 h_j^2 -14 \omega_j^2 \omega_{\tilde \jmath}^2 h_j h_{\tilde \jmath} + \omega_{\tilde \jmath}^4 h_{\tilde \jmath}^2
\end{equation}
is positive. So if $r > 0$ then the relative equilibria $\pm (h_j, 0, 0, h_{\tilde \jmath }, 0, 0)$ are stable.
\subsection{Linearisation}
Because we still don't know anything about stability if $r<0$, an alternative is to study the linearisation of the
vector field $X_{\mathcal{K}_j}$ at $\pm (h_j, 0, 0, h_{\tilde \jmath }, 0, 0)$.
Again in local coordinates, it reads \begin{equation} \label{linearisatiematrix} \nonumber
X_{\mathcal{K}_j}\left( \begin{array}{c} v_j \\ w_j \\ v_{\tilde \jmath} \\ w_{\tilde \jmath}
\end{array} \right) = \pm \frac{1}{8n} \left( \begin{array}{cccc}  0 & -\omega_j^2 h_j
& 0 & 2 \omega_j \omega_{\tilde \jmath} h_j \\ \ \omega_j^2 h_j & 0 &
2 \omega_j \omega_{\tilde \jmath} h_j & 0 \\ 0 & 2 \omega_j \omega_{\tilde \jmath} h_{\tilde \jmath} & 0 & -\omega_{\tilde \jmath}^2 h_{\tilde \jmath} \\ 2 \omega_j \omega_{\tilde \jmath} h_{\tilde \jmath} & 0 & \omega_{\tilde \jmath}^2 h_{\tilde \jmath}& 0 \end{array}
\right) \left(\begin{array}{c} v_j \\ w_j \\ v_{\tilde \jmath} \\ w_{\tilde \jmath} \end{array}
\right) + \mbox{h.o.t.} \end{equation} `h.o.t.' stands of course for
`higher order terms'. One calculates that the characteristic polynomial
of the above matrix reads $$C(\lambda) = \lambda^4 + \lambda^2 (\omega_j^4 h_j^2 + \omega_{\tilde \jmath}^4 h_{\tilde \jmath}^2 -8\omega_j^2 \omega_{\tilde \jmath}^2 h_j h_{\tilde \jmath})/(8n)^2  + 9 (\omega_j^4 \omega_{\tilde \jmath}^4 h_j^2 h_{\tilde \jmath}^2  )/(8n)^4 \ ,$$ so the
eigenvalues are the numbers $$\lambda = \pm \frac{1}{16n}\sqrt{p \pm q
\sqrt{r}} \ ,$$ where $$p := 16 \omega_j^2 \omega_{\tilde \jmath}^2 h_j h_{\tilde \jmath} -2\omega_j^4 h_j^2 -2 \omega_{\tilde \jmath}^4 h_{\tilde \jmath}^2 \ \ , \ \ q
:= 2\omega_j^2 h_j - 2\omega_{\tilde \jmath}^2 h_{\tilde \jmath} $$ and $r$ as defined previously in (\ref{r}). Note
that $C(\lambda) = C(-\lambda)$, so if $\lambda$ is an eigenvalue of (\ref{linearisatiematrix}), then so are $-\lambda, \bar
\lambda$ and $-\bar \lambda$. The reason is of course that our matrix
is conjugate to an infinitesimally symplectic matrix.\\ \indent The
next observation is that if $r \geq 0$ or $q = 0$ then $p \pm q\sqrt{r} \in
\mathbb{R}$ so the eigenvalues are purely real or purely imaginary,
dependent on the signs of $p \pm q\sqrt{r}$. On the other hand, if $r <
0$ and $q \neq 0$ then none of the eigenvalues lies on the real or the
imaginary axis. A simple analysis now leads to the following results:
\begin{itemize} 
\item If $r > 0$ then there
are four distinct purely imaginary eigenvalues.
\item If $r = 0$, we find double imaginary eigenvalues. The linearisation matrix is not semisimple. 
\item If $r < 0$ then 
none of the eigenvalues lies on the imaginary axis.  
\end{itemize} 
The set of $h_j, h_{\tilde \jmath}$ for which $r = 0$ consists of two half lines in the positive quadrant. 
 On these half lines a Hamiltonian
Hopf bifurcation \cite{vdMeer} occurs: two pairs of imaginary eigenvalues come
together and split into a quadruple of non-imaginary eigenvalues. So
the linear stability of the relative equilibria $\pm (h_j, 0, 0, h_{\tilde \jmath }, 0, 0)$  changes here from neutrally stable to unstable. The linear instability implies of course that the equilibria are unstable also in the nonlinear system.  This concludes the stability analysis of the relative equilibria. 

\subsection{Stability results} 
We summarize the results of this section in the following
\begin{corollary}\label{corollarium}
The relative equilibria $\pm (h_j, 0, 0, -h_{\tilde \jmath }, 0, 0)$  are stable. The stability of the relative equilibria $\pm (h_j, 0, 0, h_{\tilde \jmath }, 0, 0)$  depends on the bifurcation parameter $$r= \omega_j^4 h_j^2 -14 \omega_j^2 \omega_{\tilde \jmath}^2 h_j h_{\tilde \jmath} + \omega_{\tilde \jmath}^4 h_{\tilde \jmath}^2 \ .$$ They are stable for $r > 0$ and unstable for $r < 0$.
\end{corollary}

\noindent This is illustrated in the following bifurcation diagram:
\\
\begin{center} \epsfig{file=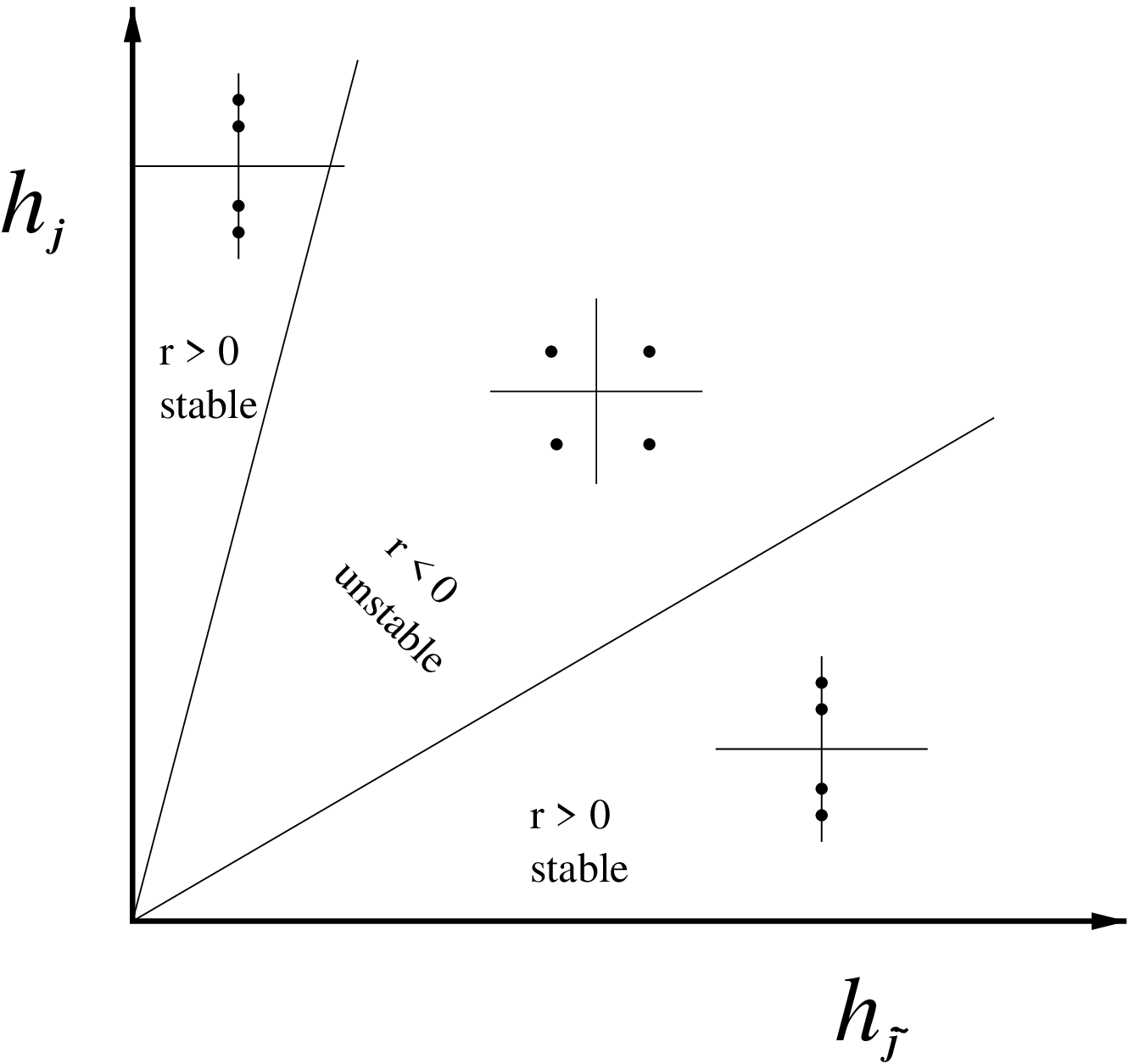 ,width=7cm ,height=6cm} \\
\end{center}

\noindent In terms of travelling wave solutions, one may interpret corollary \ref{corollarium} as follows:\\
\\
{\it \noindent The superposition of two travelling waves with wave numbers $j$ and $n/2 - j$ is stable if the two waves move in opposite directions. A superposition of waves in equal directions can be both stable and unstable. It is stable only if one of the waves is relatively small with respect to the other. Otherwise it is unstable.}

\section{Singular reduction}
\noindent In the next sections we shall use the method of singular
reduction \cite{Cushman} to understand the bifurcations of the previous paragraph geometrically, and also to show that if $r < 0$, then there are homoclinic and
heteroclinic connections between the unstable relative equilibria. \\
\\
\noindent The flow
of $X_{\mathcal{I}_j}$ induces a symplectic $S^1$-action on $S^2_{h_j} \times S^2_{h_{\tilde \jmath}}$ given by (\ref{s1actie}). The orbits of this $S^1$-action are all circles, except for
exactly the relative equilibria $(\pm h_j, 0, 0, \pm h_{\tilde \jmath }, 0, 0)$. \\
\indent The following quantities are invariant under this action:
$$\pi_j :=
u_j, \ \rho_j := u_{\tilde \jmath}, \ \sigma_j := v_j v_{\tilde \jmath} - w_j w_{\tilde \jmath}, \ \tau_j := v_j w_{\tilde \jmath} + v_{\tilde \jmath} w_j\ .$$ They satisfy the equation $$ \sigma_j^2 + \tau_j^2
= (h_j^2 - \pi_j^2)(h_{\tilde \jmath}^2 - \rho_j^2) \ \ .$$ The set $$P_{h_j,h_{\tilde \jmath}} := \{
(\pi_j, \rho_j, \sigma_j, \tau_j) \in \mathbb{R}^4 |  \sigma_j^2 + \tau_j^2
= (h_j^2 - \pi_j^2)(h_{\tilde \jmath}^2 - \rho_j^2), \ |\pi_j| \leq h_j, \ |\rho_j| \leq h_{\tilde \jmath} \}$$ therefore
constitutes the orbit space $S^2_{h_j} \times S^2_{h_{\tilde \jmath}} / _{S^1}$ Note
that we did not yet restrict ourselves to orbits of constant
$\mathcal{I}_j$: this will come later.\\ \indent Every function
on $S^2_{h_j} \times S^2_{h_{\tilde \jmath}}$ that commutes with $\mathcal{I}_j$,
reduces to a function on $P_{h_j,h_{\tilde \jmath}}$ since it is constant on orbits.
In particular, $$\mathcal{I}_j = \pi_j-\rho_j \ \mbox{and}  \ 
\mathcal{K}_j:=\frac{1}{32n} (4 \omega_j \omega_{\tilde \jmath} \sigma_j - \omega_j^2 \pi_j^2 -
\omega_{\tilde \jmath}^2 \rho_j^2 ) \ .$$
\noindent The reduction map is the map $$ \mathcal{F}^{(3)}: (u_j, v_j, w_j, u_{\tilde \jmath}, v_{\tilde \jmath}, w_{\tilde \jmath})
\mapsto (\pi_j, \rho_j, \sigma_j, \tau_j)$$ which goes from $S^2_{h_j} \times S^2_{h_{\tilde \jmath}}$ to
$P_{h_j,h_{\tilde \jmath}}$. The reduction map is a submersion everywhere, except of
course at the relative equilibria. Unfortunately it is not possible yet
to make a drawing of $P_{h_j,h_{\tilde \jmath}}$ since it can not be embedded in
$\mathbb{R}^3$. But there is an elegant way to overcome this problem.
\\ \indent One sees that both $\mathcal{I}_j$ and $\mathcal{K}_j$ are
independent of $\tau_j$, that is these Hamiltonians are invariant under
the $\mathbb{Z}_2$-action generated by the mapping $\tau_j \mapsto
-\tau_j$. This symmetry is induced by the discrete symmetries of the original
Hamiltonian (\ref{hamfpu}), cf. \cite{Rink2}. The orbits of the
$\mathbb{Z}_2$-action consist of one point if $\tau_j = 0$ and otherwise
of two points. One can reduce the $\mathbb{Z}_2$-action by simply
forgetting about $\tau_j$: the reduction map is $(\pi_j, \rho_j, \sigma_j, \tau_j)
\mapsto (\pi_j, \rho_j, \sigma_j)$. The reduced space is the set $$P_{h_j,
h_{\tilde \jmath}}/_{\mathbb{Z}_2} = \{(\pi_j, \rho_j, \sigma_j) \in
\mathbb{R}^3 | \sigma_j^2 \leq (h_j^2 - \pi_j^2)(h_{\tilde \jmath}^2 - \rho_j^2),
\ |\pi_j| \leq h_j, \ |\rho_j| \leq h_{\tilde \jmath} \} \ .$$ Below, we draw $P_{h_j,
h_{\tilde \jmath}}/_{\mathbb{Z}_2}$ for $h_j, h_{\tilde \jmath} > 0$ :  
\begin{center} \epsfig{file=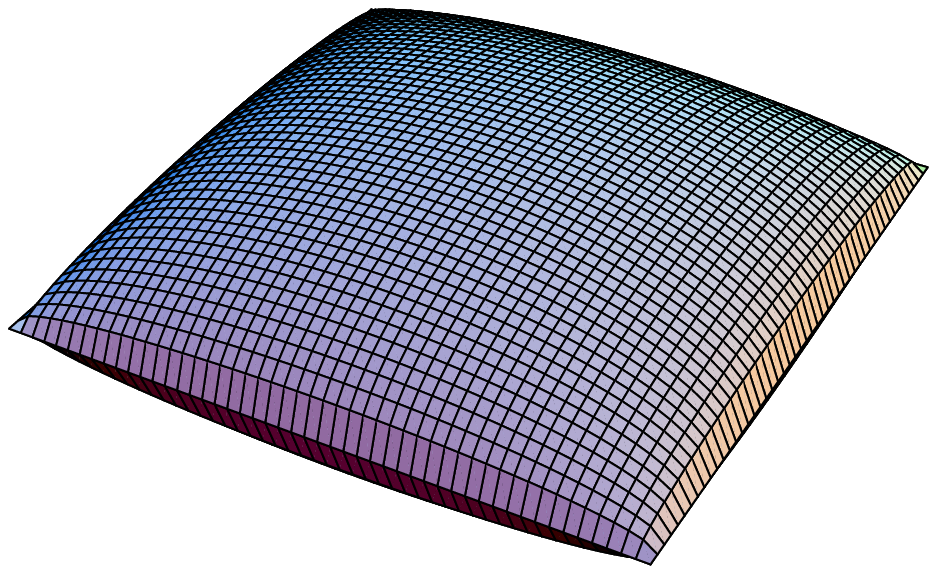
,width=6.5cm ,height=5cm} \\ 
\end{center}
\noindent $P_{h_j,h_{\tilde \jmath}}/_{\mathbb{Z}_2} = S^2_{h_j} \times S^2_{h_{\tilde \jmath}} / _{S^1 \times \mathbb{Z}_2}$ has the shape of a solid
pillow. The corners of the
pillow are cone-like singularities that represent the relative equilibria $(\pm h_j, 0, 0, \pm h_{\tilde \jmath }, 0, 0)$.\\
\indent The level sets of $\mathcal{I}_j = \pi_j - \rho_j$ are two dimensional planes. The intersection of such a plane with the pillow is a topological disk, a point or empty. Near the singularities $(\pi_j, \rho_j, \sigma_j) = \pm (h_j, -h_{\tilde \jmath}, 0)$ the disks are very small, indicating that the relative equilibria $\pm (h_j, 0, 0, -h_{\tilde \jmath }, 0, 0)$ are stable. But near the other two corners of the pillow, the singularities $\pm (h_j, h_{\tilde \jmath}, 0)$, the level set of $\mathcal{I}_j$ intersects the pillow in a very large set:
\begin{center}
\epsfig{file=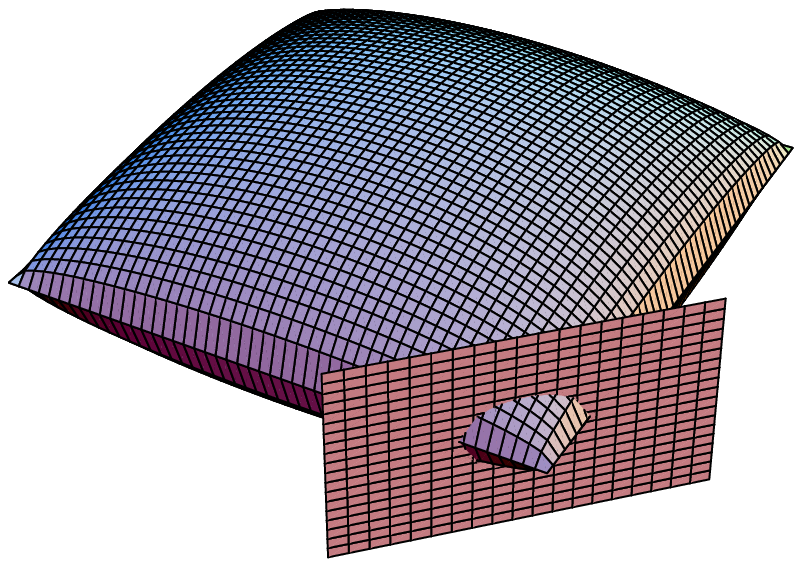, width=6cm, height=4.5cm} \hspace{1cm}
\epsfig{file=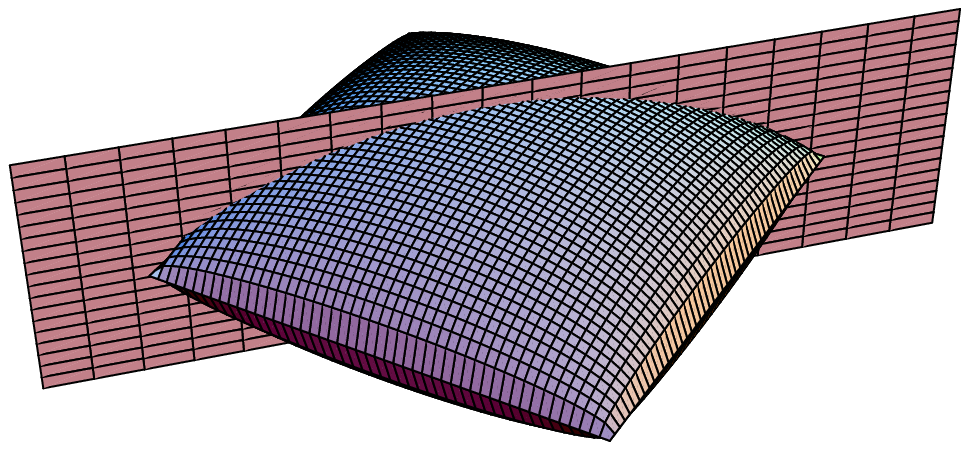, width=6cm, height=4.5cm}
\end{center}
\hskip 2.2cm {\footnotesize $\mathcal{I}_j$ near $h_j + h_{\tilde \jmath}$ \hskip 4.8cm $\mathcal{I}_j$ near $h_j - h_{\tilde \jmath}$ }\\
\\
\noindent Let us consider the level set of $\mathcal{I}_j$ that passes exactly through the singularity $(h_j, h_{\tilde \jmath}, 0)$. It is the plane $\mathcal{I}_j = \pi_j - \rho_j = h_j - h_{\tilde \jmath}$. The intersection of this plane with the pillow is the topolocal disk $$\{ (\pi_j, \rho_j, \sigma_j) \in \mathbb{R}^3 | \sigma_j^2 \leq (h_j^2 - \pi_j^2)(h_{\tilde \jmath}^2 - (\pi_j + h_{\tilde \jmath} - h_j)^2) 
\ , \rho_j = \pi_j + h_{\tilde \jmath} - h_j\}\ .$$ If $h_j = h_{\tilde \jmath}$, then this disk has two singular points. It has one singular point if $h_j \neq h_{\tilde \jmath}$. The intersection of a level set of $\mathcal{K}_j$ with this plane is a parabola. The parabola that contains the singularity $(h_j, h_{\tilde \jmath}, 0)$ is given by the formulas $$\sigma_j = a(\pi_j):= 
\frac{1}{4\omega_j\omega_{\tilde \jmath}}(\omega_j^2 \pi_j^2 + \omega_{\tilde \jmath}^2 ( \pi_j + h_{\tilde \jmath} - h_j)^2 -\omega_j^2 h_j^2 - \omega_{\tilde \jmath}^2 h_{\tilde \jmath}^2)\ , \ \rho_j = \pi_j + h_{\tilde \jmath} - h_j\ . $$ We now make a linear approximation to both this parabola and the singular disk at the singular point $(h_1, h_2, 0)$. So we calculate the derivative $\frac{d a}{d \pi_j}\left|_{\pi_j = h_j} \right. = \frac{1}{2 \omega_j \omega_{\tilde \jmath}}(\omega_j^2 h_j + \omega_{\tilde \jmath}^2 h_{\tilde \jmath})$. On the other hand, the cone-like singularity of the singular disk is approximated by $$\{ |\sigma_j| \leq 2 \sqrt{h_j h_{\tilde \jmath}}(h_j-\pi_j), \pi_j \leq h_j, \rho_j = \pi_j + h_{\tilde \jmath} - h_j\}\ .$$ So the tangent line to the parabola points into the cone exactly if $-2\sqrt{h_j h_{\tilde \jmath}} < \frac{1}{2 \omega_j \omega_{\tilde \jmath}}(\omega_j^2 h_j + \omega_{\tilde \jmath}^2 h_{\tilde \jmath}) <  2\sqrt{h_j h_{\tilde \jmath}}$, that is if $r < 0$. In this case, the critical point $(\pi_j, \rho_j, \sigma_j) = (h_j, h_{\tilde \jmath}, 0)$ is clearly unstable, which agrees with our previous analysis. The tangent to the parabola does not point into the cone if $r > 0$. The following two pictures represent the two possibilities: \\
\begin{center}
\epsfig{file=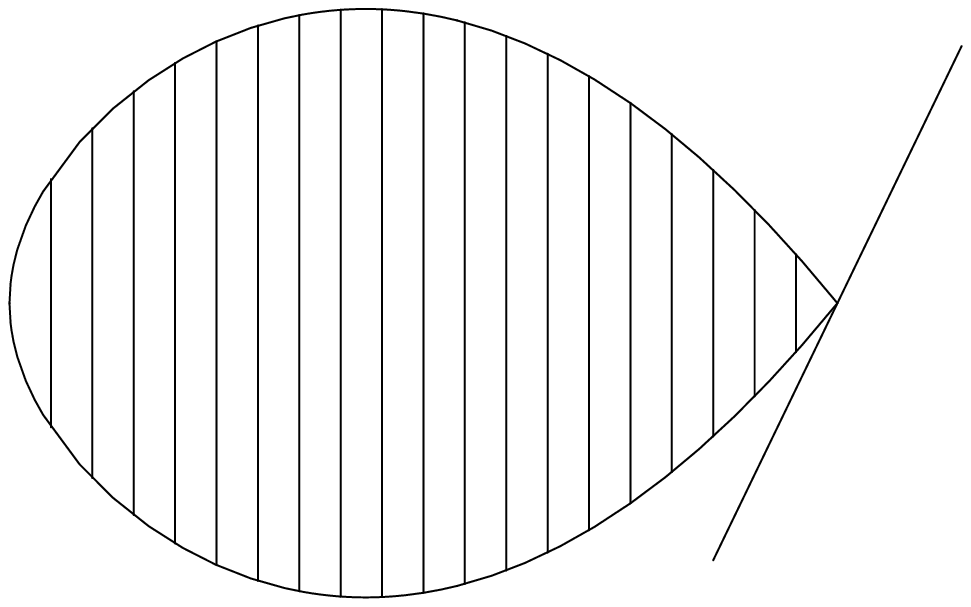, width=5cm, height=3cm} \hspace{1cm}
\epsfig{file=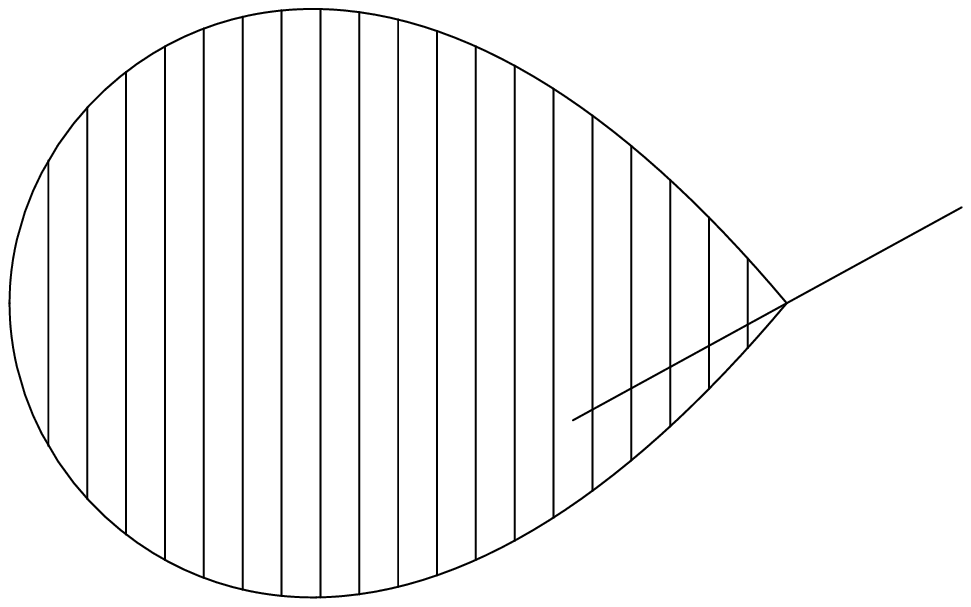, width=5cm, height=3cm}
\end{center}
\hskip 3.1cm {\footnotesize $r > 0$ \hskip 5.3cm $r < 0$}\\
\\
\noindent In the case that $r > 0$, we now see that $\mathcal{K}_j$, restricted to $(P_{h_j,h_{\tilde \jmath}}/_{\mathbb{Z}_2}) \cap \mathcal{I}_j^{-1}(h_j-h_{\tilde \jmath})$, is extremal at $(h_j, h_{\tilde \jmath}, 0)$. In other words, $(h_j, 0, 0, h_{\tilde \jmath }, 0, 0)$ is a stable relative equilibrium inside the singular fiber $(S^2_{h_j} \times S^2_{h_{\tilde \jmath}} ) \cap \mathcal{I}_j^{-1}(h_j-h_{\tilde \jmath})$. A theorem of Montaldi \cite{Montaldi} then states that $(h_j, 0, 0, h_{\tilde \jmath }, 0, 0)$ is stable in $S^2_{h_j} \times S^2_{h_{\tilde \jmath}}$. This is an entirely geometric way to see the Hamiltonian Hopf bifurcation and the stability change of the relative equilibrium $(h_j, 0, 0, h_{\tilde \jmath }, 0, 0)$. The same argument holds for the equilibrium $(-h_j, 0, 0, -h_{\tilde \jmath }, 0, 0)$.

\section{Pinched tori and monodromy}  
\noindent We shall argue that the reduced system on $S^2_{h_j} \times
S^2_{h_{\tilde \jmath}}$ induced by $\mathcal{K}_j$ has nontrivial monodromy and that there are homoclinic and heteroclinic connections between the relative equilibria $\pm(h_j, 0, 0, h_{\tilde \jmath }, 0, 0)$ if $r<0$.\\ 
\indent This can all be shown by simply drawing a picture of $(P_{h_j,h_{\tilde \jmath}}/_{\mathbb{Z}_2}) \cap \mathcal{I}_j^{-1}( h_j-h_{\tilde \jmath} )$ and its foliation into the level sets of $\mathcal{K}_j$. Recall that $(P_{h_j,h_{\tilde \jmath}}/_{\mathbb{Z}_2}) \cap \mathcal{I}_j^{-1}( h_j-h_{\tilde \jmath} )$ is a topological disk that contains one singular point if $h_j\neq h_{\tilde \jmath}$ and two singular points if $h_j=h_{\tilde \jmath}$. Each point inside the disk represents two three-dimensional tori in $T^*\mathbb{R}^4$. The regular points on the boundary of the disk each represent one three dimensional torus. The singular points, which also lie on the boundary of the disk, represent a `singular' two-dimensional torus. This singular torus has the interpretation of a superposition of travelling waves in the FPU chain.\\
\indent Let us first consider the case that $h_j = h_{\tilde \jmath}$. In that case the reduced space is bounded by parabolas
$$(P_{h_j,h_{j}}/_{\mathbb{Z}_2}) \cap \mathcal{I}_j^{-1}( 0 )  = \{ (\pi_j, \rho_j, \sigma_j) \in \mathbb{R}^3 | \ |\sigma_j| \leq h_j^2 - \pi_j^2\ ,\ \rho_j = \pi_j\}$$
It has two singular points which lie on the same level set of $\mathcal{K}_j$. This level set is also a parabola. Furthermore, note that $r = (\omega_j^4 + \omega_{\tilde \jmath}^4 - 14 \omega_j^2 \omega_{\tilde \jmath}^2) h_j^2 = 16 (1- \omega_j^2 \omega_{\tilde \jmath}^2)h_j^2$, which is negative if and only if $\omega_j \omega_{\tilde \jmath} > 1$ if and only if $n/12 < j < n/4$. But then the reduced space simply looks like this:
\begin{center}
\epsfig{file=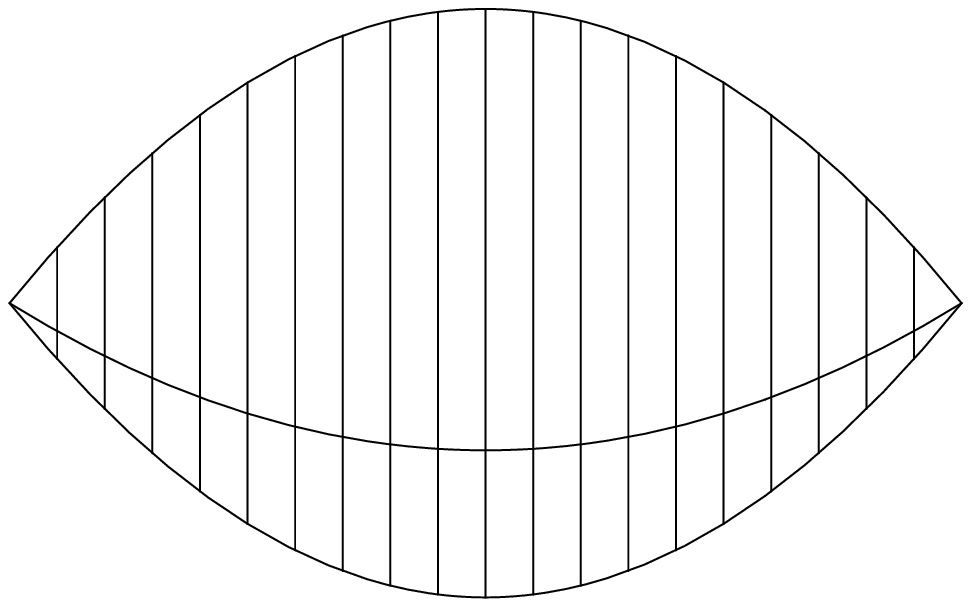, width=5cm, height=3cm} \hspace{1cm}
\end{center}
We observe that there is a heteroclinic connection between the two singular points. In the reconstructed system on $S^2_{h_j} \times S^2_{h_{\tilde \jmath}}$ this corresponds to a doubly pinched torus: two focus-focus singular points of which the stable and unstable manifolds coincide. In the original phase space $T^*\mathbb{R}^{4}$, this doubly pinched torus is again reconstructed as a heteroclinic connection between two-dimensional tori. They are connected by their `whiskers' which have dimension four and are both diffeomorphic to $\mathbb{R} \times T^3$. \\
\indent The heteroclinic connection has the following interpretation. If one starts with a motion that is nearly the superposition of two travelling waves with equal energy and in equal direction, then after a certain `incubation period' both travelling waves will come to a halt and turn around until the motion looks very much like the superposition of two travelling waves the directions and energies of which are again equal, although the direction is opposite to the direction in the beginning. This process continues and the superposed waves keep changing direction.\\ 
\indent In the case that $h_j \neq h_{\tilde \jmath}$ and $r < 0$, the singular level set generically looks like this:
\begin{center}
\epsfig{file=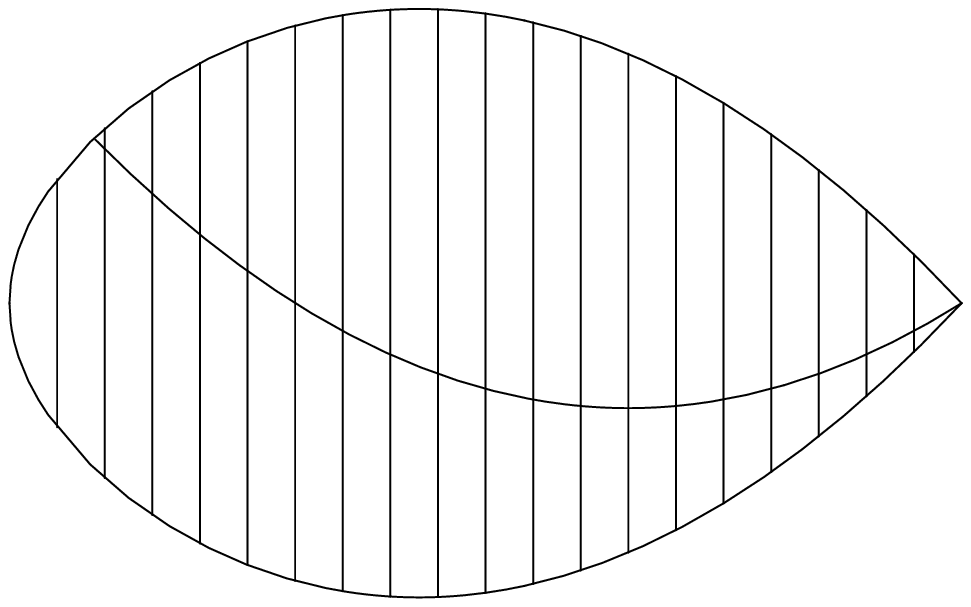, width=5cm, height=3cm} \hspace{1cm}
\end{center}
We see that there is a homoclinic connection that connects the singular point to itself. In the reconstructed system on $S^2_{h_j} \times S^2_{h_{\tilde \jmath}}$ this corresponds to a singly pinched torus: the stable and unstable manifold of the focus-focus singular point coincide. In  $T^*\mathbb{R}^{4}$ the pinched torus is a homoclinic connection of a two-dimensional torus to itself. The two-dimensional torus again represents the superposition of travelling waves in the same direction with wave number $j$ and $\tilde \jmath$, which now do not have equal energy.\\
\indent This completely describes the interaction between travelling waves with wave number $j$ and travelling waves with wave number $n/2 -j$. The travelling waves do not interchange any energy. Still, their influence on one another is such that each of the travelling waves drastically changes its momentum and thus its direction. 
\subsection{Monodromy}
\noindent It was shown in \cite{Tienzung} that the presence of a pinched torus implies monodromy: the regular Liouville tori in  $S^2_{h_j} \times
S^2_{h_{\tilde \jmath}}$ do not constitute a trivial torus bundle. Instead, there is monodromy and the monodromy map is known. When we reconstruct, we see that the regular tori in $T^*\mathbb{R}^{4}$ and $T^*\mathbb{R}^{n-1}$ can not form a trivial bundle either. Nontrivial monodromy is an important obstruction to the existence of global action-angle variables. Recall that the pinched torus is present if $n \geq 6$ is even. We conclude that if $n \geq 6$ is even, then the integrable normal form (\ref{normaalvorm}) does not admit global action-angle variables. \\
\indent We have now described how the level sets of the integrals of the Birkhoff normal form (\ref{normaalvorm}) globally foliate the phase space. The same foliation describes all the integrable structure that is present in the low energy domain of the original FPU Hamiltonian (\ref{hamfpu}). In a certain sense, the KAM tori in the low energy domain of the phase space of (\ref{hamfpu}) will also form a nontrivial bundle.

\section{Numerical solutions for 16 and 32 particles}
I tried to detect the direction reversing waves numerically in the original periodic FPU system (\ref{hamfpu}). This system is not integrable, but in the low energy domain it is approximated by the integrable normal form. Therefore we expect that the low energy solutions of FPU Hamiltonian (\ref{hamfpu}) behave as predicted by the Birkhoff normal form. \\
\indent Let us study the periodic FPU chain with $n=16$ particles, a number chosen by Fermi, Pasta and Ulam themselves. We shall investigate solutions that start out as a superposition of two travelling waves with wave numbers $j = 3$ and $\tilde \jmath = \frac{16}{2} - 3 = 5$. Note that  $\frac{n}{12} = \frac{4}{3} < j = 3 < \frac{n}{4} = 4$, so if we give both waves equal energy and equal direction, then we expect the solution to reverse its direction. \\
\indent  Let us take the initial conditions $\bar q (0) = (0,0,.25,0,.25,0,0,0,0,0,0,0,0,0,0)$ and $\bar p (0) = (0,0,0,0,0,0,0,0,0,0,.25,0,.25,0,0)$, such that $\mathcal{H}_3 (0) =  \mathcal{H}_5 (0) = u_3(0) = u_5(0) = .0625$. The total energy is $H = 0.1742$. The angular momenta $u_3$ and $u_5$ measure the direction of the waves. Their values in the course of time are plotted here:
\begin{center}  %begincondities in file init161
\epsfig{file=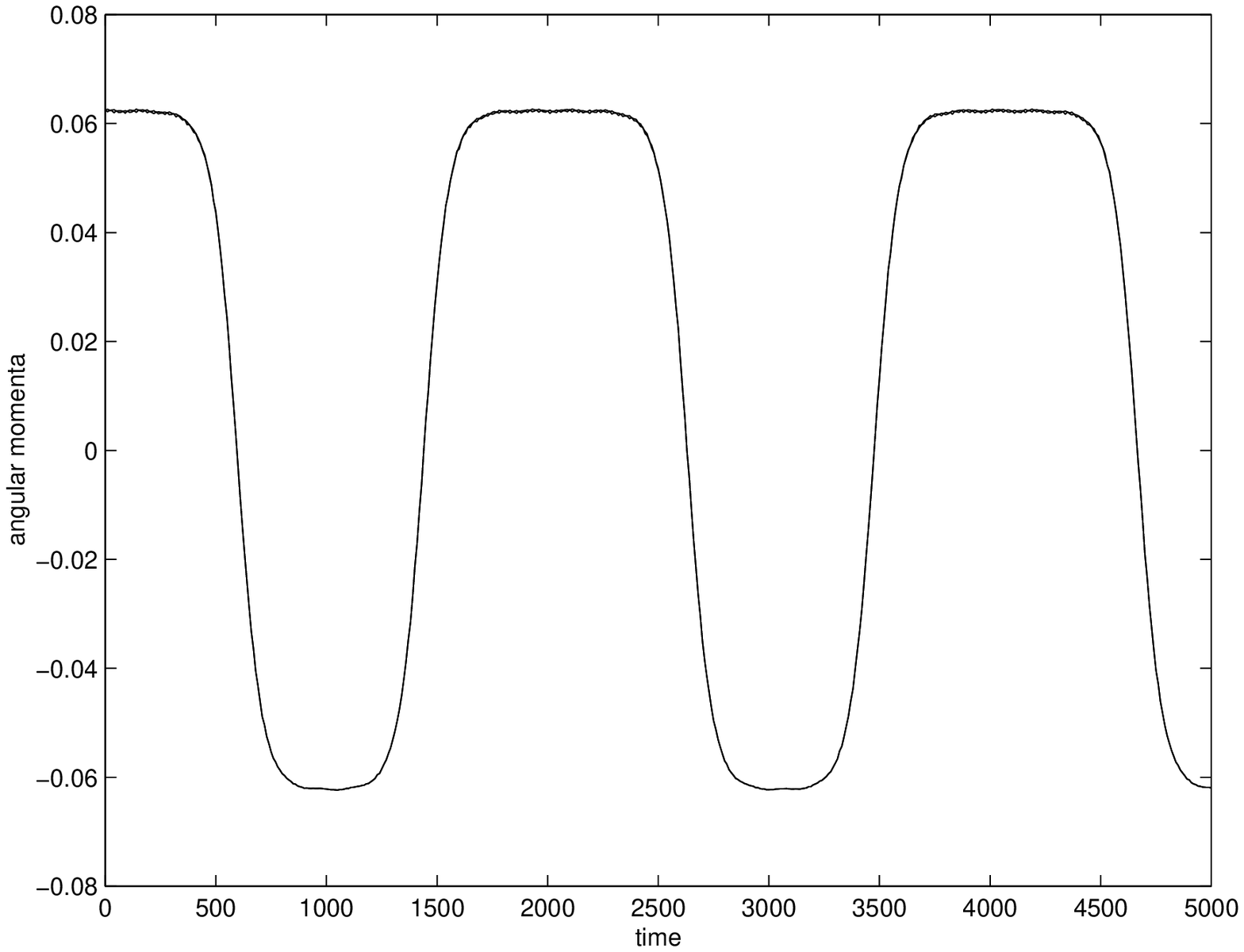, width=8cm, height=5cm} \hspace{1cm}
\\
\footnotesize{Superposed travelling waves with wave numbers $3$ and $5$ in the FPU chain with $16$ particles interact as predicted by the normal form: the angular momenta exhibit a relaxation oscillation between their maximal and minimal values.}
\end{center}
\noindent Note that this line is double: the angular momenta remain equal all the time, just as was predicted by the normal form. Furthermore, they vary between their maximal and minimal values at the given energy. The heteroclinic connection of the normal form is clearly visible here: the angular momenta seem to stick to their maximal and minimal value for quite some time, before moving off again. It is not very likely though that this heteroclinic connection really exists in the original system. Generically, coinciding stable and unstable manifolds start having transverse intersections under small perturbations, resulting in chaotic orbits. On the other hand, the angle under which they intersect is extremely small, such that this chaos is hardly visible. Hence our observations do not look chaotic at all. \\
\indent To compare, I also investigated how waves with wave numbers $2$ and $5$ interact. Note that $\tilde 2 \neq 5$. So I took $\bar q (0) = (0,.25,0,0,.25,0,0,0,0,0,0,0,0,0,0)$ and $\bar p (0) = (0,0,0,0,0,0,0,0,0,0,.25,0,0,.25,0)$, such that $\mathcal{H}_2 (0) =  \mathcal{H}_5 (0) = u_2(0) = u_5(0) = .0625$. The total energy is $H = 0.1528$. The values of $u_2$ and $u_5$ are depicted here on the same time scale:
\begin{center}     % begincondities in file init162
\epsfig{file=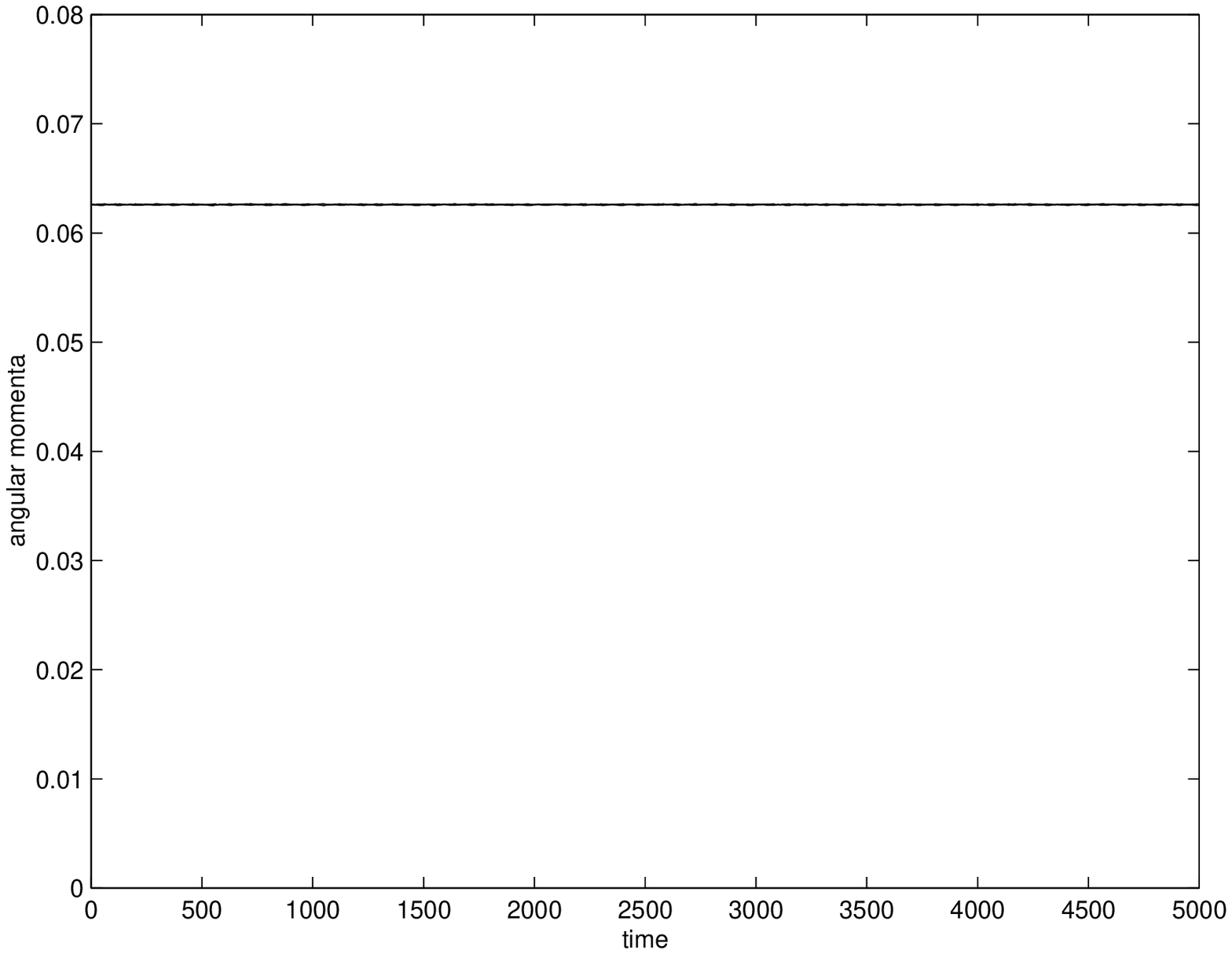, width=6cm, height=4cm} \hspace{1cm}\\
\footnotesize{As predicted, waves with wave numbers $2$ and $5$ do not interact.}
\end{center}
As predicted by the normal form, waves with wave numbers $2$ and $5$ do not interact at all.\\
\indent I also investigated how the waves with wave numbers $3$ and $5$ interact at a higher energy level. So we start with $\bar q (0) = (0,0,.5,0,.5,0,0,0,0,0,0,0,0,0,0)$ and $\bar p (0) = (0,0,0,0,0,0,0,0,0,0,.5,0,.5,0,0)$, such that $\mathcal{H}_2 (0) =  \mathcal{H}_5 (0) = u_2(0) = u_5(0) = .25$ and the total energy is $H = 0.7048$. The result:
\begin{center}   % begincondities in file init163
\epsfig{file=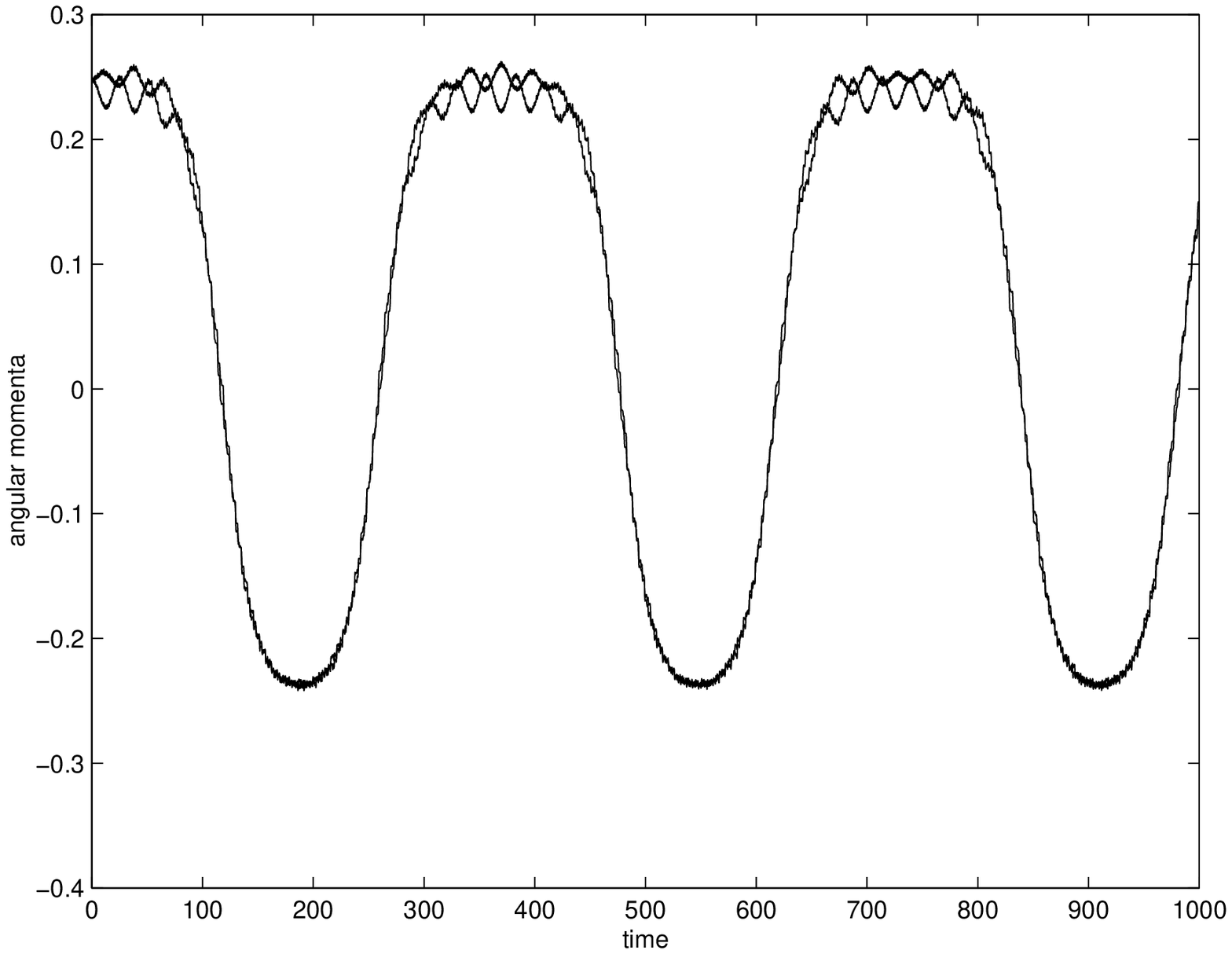, width=6cm, height=4cm} \hspace{1cm}\\
\footnotesize{Also at a higher energy level, the normal form is a good approximation.}
\end{center}
Note that even at this high energy level the normal form still constitutes a very good approximation! This is a remarkable observation, but similar phenomena have been observed in other systems.\\
\\
\noindent Finally, I integrated the chain with $32$ particles with initial conditions $\bar q_j (0) = \bar p_j (0) = 0$ for all $j$ except $\bar q_7(0) = \bar q_9(0) = \bar p_{23}(0) = \bar p_{25}(0) = .25$ such that $\mathcal{H}_7(0) = \mathcal{H}_9(0) = u_7(0) = u_9(0) = .0625$ and $H = 0.1762$. The results are still very nice:
\begin{center}   % begincondities in file init321
\epsfig{file=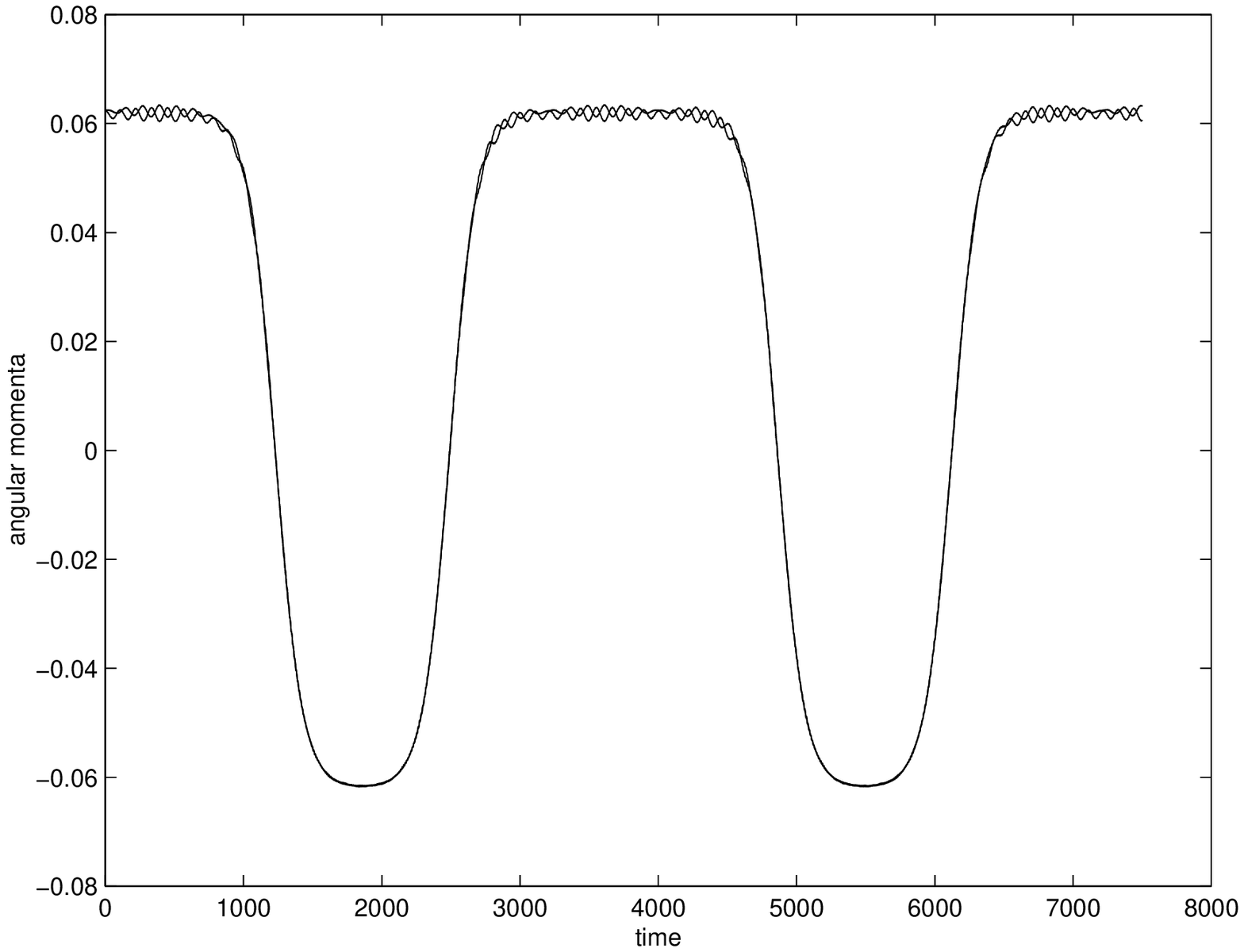, width=6cm, height=4cm} \hspace{1cm}\\
\footnotesize{For the chain with $32$ particles, the normal form works perfectly at this energy level.}
\end{center}
But at a higher energy the normal form is no longer a good approximation when $n = 32$. If one chooses the initial conditions $\bar q_j (0) = \bar p_j (0) = 0$ for all $j$ except $\bar q_7(0) = \bar q_9(0) = \bar p_{23}(0) = \bar p_{25}(0) = .5$, then $\mathcal{H}_7(0) = \mathcal{H}_9(0) = u_7(0) = u_9(0) = .25$ and $H = 0.7091$. The angular momenta interact as follows:
\begin{center}   % begincondities in file init322
\epsfig{file=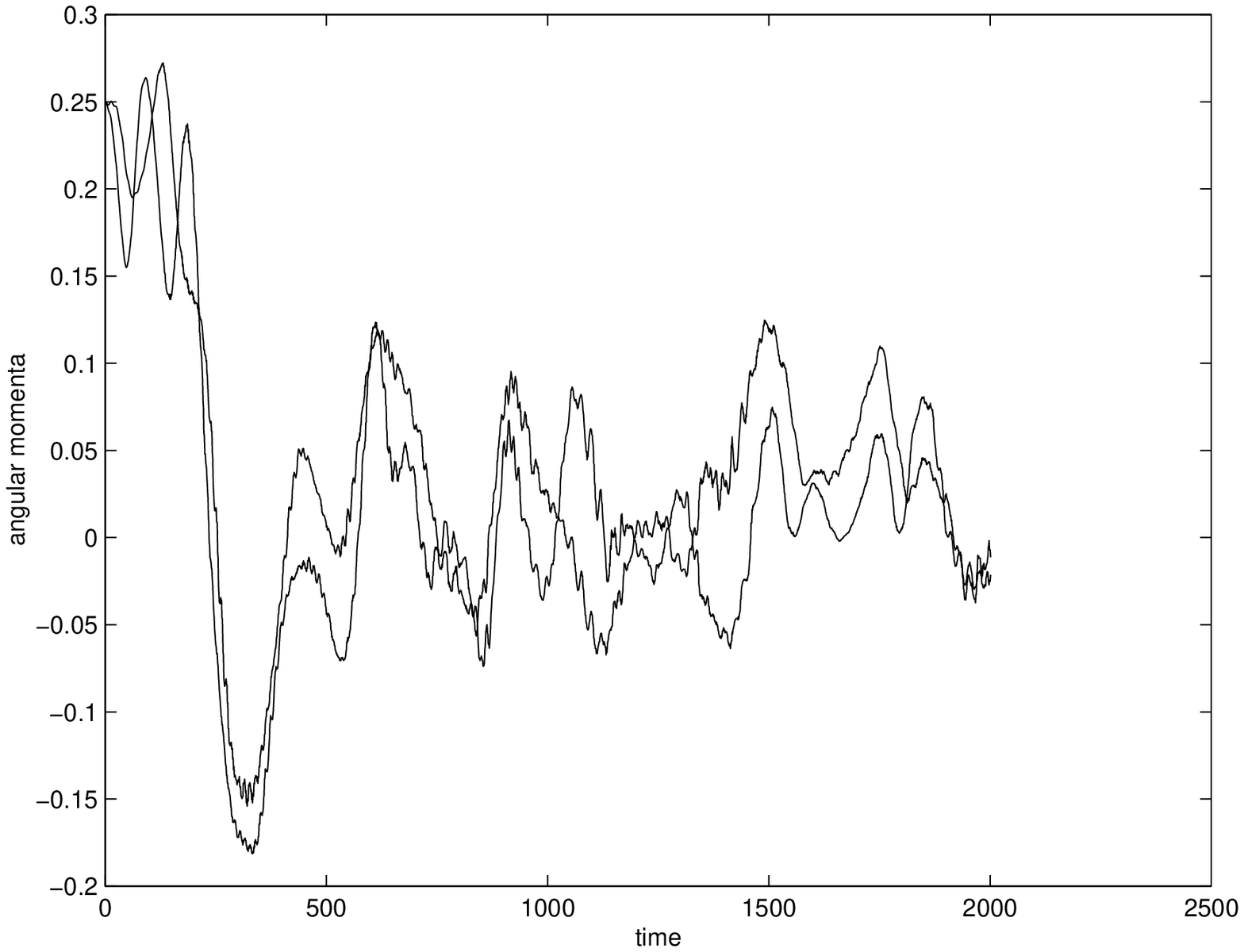, width=6cm, height=4cm} \hspace{1cm}\\
\footnotesize{At high energy, the normal form does not make much sense anymore  for $n=32$.}
\end{center}
Thus, the domain of validity of the normal form shrinks if $n$ grows. This is to be expexted, because when $n$ is very large, near-resonances start spoiling the validity of the normal form. Still, the normal form gives us some surprisingly good predictions, even at energies where one would not expect this anymore.

\section{Conclusions and discussion}
We considered the Birkhoff normal form of the periodic Fermi-Pasta-Ulam chain with periodic boundary conditions and quartic nonlinearities. In the domain of the phase space where the solutions have low energy, this Birkhoff normal form constitutes an approximation of the original Fermi-Pasta-Ulam chain. And because of symmetries and nonresonance properties, it is Liouville integrable, see \cite{Rink2}. We studied how the level sets of the integrals foliate the phase space, because one expects that a lot of the integrable structure of the normal form is still present in the low energy domain of the original FPU chain. \\ 
\indent The most interesting phenomena happen when the number of particles $n$ is even. In that case the integrals are quadratic
 and quartic functions of the phase space variables. Using the purely geometric methods of regular and singular reduction \cite{Cushman}, we showed that the foliation has singular elements, pinched tori.\\
\indent It is well-known \cite{Duistermaat}, \cite{Tienzung} that pinched
tori imply monodromy: the Liouville tori do not form a trivial torus bundle over the set of regular values of the integrals. This is important information about how the Liouville tori are glued together globally, for instance on an energy level set. Among others, monodromy is an important topological obstruction to the
existence of global action-angle variables. Because the Birkhoff normal form approximates the FPU chain especially well in the low energy domain, we expect that the KAM tori on the low energy level sets are `glued together' similarly.\\
\indent At the same time, our study unravels interesting dynamical information. We are able to determine how waves with different wave numbers interact. It turns out that waves with wave number $j$ can only interact with waves of which the wave number is $n/2 -j$. And even though these waves do not interchange any energy, their interaction is far from trivial. The pinched tori that were mentioned before are homoclinic and heteroclinic connections between solutions which are a superposition of travelling waves with these wave numbers. Thus it can happen that these superposed travelling waves change their direction.\\
\indent In
the original FPU chain (\ref{hamfpu}), we indeed find these
direction reversing travelling waves numerically in the chains with $16$ and $32$ particles. They form a class of interesting new solutions of the periodic FPU chain. \\
\indent Surprisingly, one can even observe direction reversing travelling waves at rather high energy levels, where the Birkhoff normal form is a very questionnable approximation. It would be interesting to study how robust the Liouville tori near a pinched torus are under Hamiltonian perturbations. Maybe the KAM theorem can produce extra strong conclusions for systems with monodromy, which enables us to understand the validity of the normal form at high energy. \\
\indent Finally, the reader should be aware of other wave reversal phenomena that have been observed in the literature. I especially refer to \cite{Degasperis} which studies the {\it `boomeron'}, a soliton that comes back. Thus, we have found yet another interesting link between the Fermi-Pasta-Ulam chain and integrable wave equations.

\section{Acknowledgement}
The author thanks Richard Cushman and Ferdinand Verhulst for many valuable discussions and remarks. Odo Diekmann, Hans Duistermaat, Darryl Holm and Claudia Wulff gave some nice comments and references.

\end{document}